\documentclass[twocolumn, secnumarabic, amssymb, noshowpacs,nobibnotes, aps, showpacs,prb]{revtex4-1}
\usepackage{graphics}
\usepackage[dvips]{graphicx}
\usepackage{dcolumn}
\usepackage{amsmath}
\usepackage{lipsum}

\begin{document}
\title{Improved optical transitions theory for superlattices and periodic systems; new selection rules}
\author{Pedro Pereyra}
\address{F\'{i}sica Te\'{o}rica y Materia Condensada, UAM-Azcapotzalco, Av. S. Pablo 180, C.P. 02200, M\'{e}xico D. F., M\'{e}xico }
\date{\today}

\begin{abstract}
Using the genuine superlattice eigenvalues
and  eigenfunctions, and the eigenfunctions parity symmetries, discussed recently,  new optical-transition selection rules are derived, and a nobel and basically different theoretical approach for explicit calculation of optical responses of light-emitting periodic structures is presented here. To show the scope of this approach, we review and revisit a number of photoluminescence and infrared measurements reported in the literature. The photoluminescence and infrared spectra of superlattices based on  (Al,Ga)As and (In,Ga)N, with clear spectrum features, high resolution and different superlattice characteristics, comprising small ($\sim $10) and large ($\sim $400) number of unit-cells, wide and narrow barrier and valley widths, varying from 2.5nm to $\sim $40nm, have been recalculated. The plots obtained here for the optical response of the chosen systems,  reproduce rather well the observed
photoluminescence or infrared spectra. We show that the narrow peaks clustered in groups that were observed in blue-emitting superlattices, but couldn't be explained before, are faithfully reproduced and fully understood. We show that this replication of groups of peaks and the isolated peak  observed in high resolution spectra, are interesting effects neatly determined by the recently unveiled tunable surface-energy-levels detachment.
Among the various properties and differences discussed in
the paper, we find significant that the observed optical transitions, forbidden before, are now allowed and accounted. Since the number of allowed matrix-elements,  $\sim n^2$ $n_cn_v/2$, can be extremely large 
when the number of unit cells, $n$, and the number of subbands in the conduction and valence bands, $n_c$ and $n_v$, are large, we devote the last part of this paper to reduce this number and
to show that, essentially, the same spectra is obtained when,
besides the symmetry selection rules, other leading order selection rules, closely
related to intra-subband symmetry, are introduced. These rules
reduce the number of matrix-elements evaluations from $\sim n^2$$n_cn_v/2$
to $\sim n$$n_cn_v/2$, i.e., depending on the SL, from about 1000
to 100. We comment also on a third rule, that picks up the
contributions of the surface and edge states, and show that it
reduces further the number of transitions to $N_s\leq n_cn_v$.
With these rules, the main peaks are conserved and their number
practically matches with that of the actual spectrum. Excellent
agreements with experimental results are found.
\end{abstract}

\maketitle

\section{Introduction}
Despite the broad theoretical and empirical  knowledge of the
light-matter interactions\cite{Band} and the overwhelming variety
of device applications, the actual quantum mechanical description
of  photoemission and photoabsorption processes involving
periodic semiconductor structures, suffer from important limitations. The
main problem in the theoretical calculations, using the golden
rule
\begin{eqnarray}
|\langle\psi_{\rm f} |H_{\rm int}|\psi_{\rm i}\rangle |^2/[E_{\rm f}-E_{\rm i}+\hbar \omega)^2+\Gamma_{\rm i}^2],
\end{eqnarray}
has been the lack of explicit knowledge of the initial and final
states, $|\psi_{i}\rangle$ and $|\psi_{f}\rangle $, and of the
corresponding energies, $E_{i}$ and $E_{f}$.  In the standard
approach to periodic systems, the energy levels become  bands or
subbands, and  the initial and final superlattice (SL) states, written in terms of
Bloch functions,\cite{Bloch}  are generally unknown.\cite{Bastard}
However, in the alternative theory of finite periodic systems, the identifiable 
energy levels are recovered and the quantum states of
semiconductor heterostructures, like quantum wells and
superlattices,  are explicitly known.\cite{Pereyra2005}  Our
purpose here is to present an improved optical transitions theory
for periodic systems, based on {\it bona fide} eigenfunctions and
their parity symmetries recently unravelled.\cite{Pereyra1}  

To understand better the scope and the need of an alternative approach to calculate the optical response, let as briefly outline, in this introduction, the essential features, limitations and advantages of the standard and the new approach for the calculation of the physical quantities relevant to the light emission processes.

\begin{figure*}
\begin{center}
\includegraphics[width=440pt]{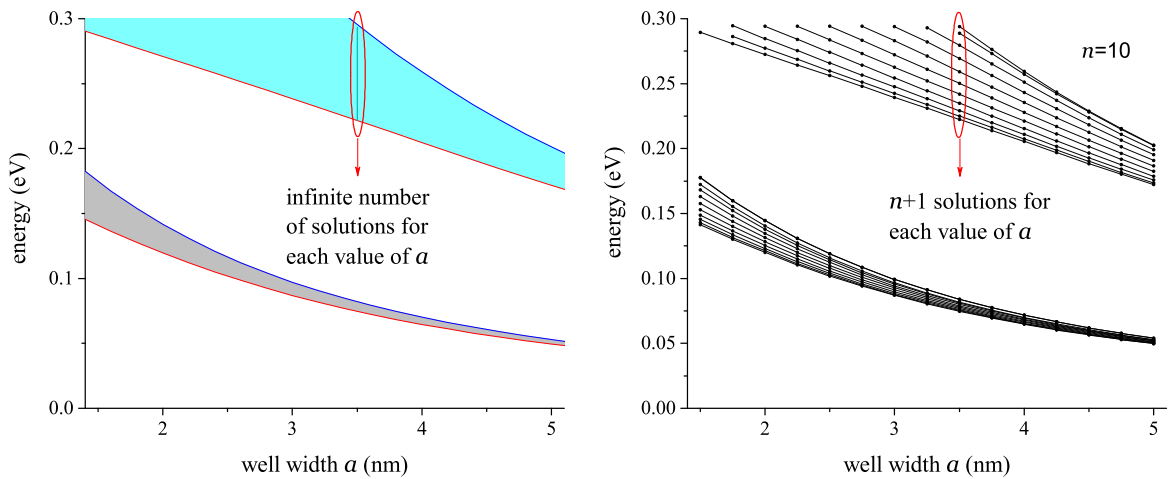}
\caption{Continuous and discrete first two subbands of the conduction band, as functions of the well width $a$, obtained for a Koronig-Penney-like periodic potential in the standard approach (left) and in the TFPS (right).}
\end{center}\label{FigCD1}
\end{figure*}

Soon after the introduction of semiconductor
superlattices,\cite{Keldysh1962,EsakiTsu1970}   the miniband
structures of direct and indirect band gap semiconductors were
experimentally and theoretically
confirmed,\cite{Esaki1972,Chang1974,Dingle1974,Mukherji1975,Miller1976,Chang1977,
SaiHalaszChang1978,Chemla1982,Belle1985,Capasso1986,Helm1989,LuoFurdyna1990,
Luscombe1991,Rauch1997,Capasso1997,Petrov1997,Heer1998} and the optical
properties of  superlattices became overwhelmingly
studied.\cite{Dingle1975,TsuKoma1975,Chomete1986,Yuh1987,Fu1989,Helm1991,Fu1995,Chemla1996,
Haug1997,Leo1998} Concerning the theoretical approaches applied to study these systems, Leo Esaki noticed that whereas in
reality the SLs contain a finite number of layers, with a finite
number of atomic cells each, the standard  theoretical approaches
tacitly assume that the SLs  are infinite-periodic structures with
alternating layers containing also an infinite number of atomic
cells.\cite{EsakiLesHuches} In fact, the wave functions are generally written
as $\psi ({\bf r})=\sum_l u_{n_{l}}({\bf r})f_{l}({\bf r})$,  with
$u_{n_{ l}}({\bf r})$ the periodic part of the host-semiconductor
Bloch's function at band $n_{l}$, and $f_{l}({\bf r})\propto\exp[{i
{\bf k}_{\bot}\cdot {\bf r}_{\bot}}]\chi_{l}(z)$ the envelope wave
function. For SLs this is, again, written in terms of Bloch-type
functions $\chi_{\mu}(z)$ $=\exp(iqz)u_{\mu}(z)$, characterized by
a subband index $\mu$ and a continuous wave number $q$ that is
then artificially discretized, via the cyclic boundary condition.
But, as Bastard states in Ref. [\onlinecite{Bastard}], ``...the
Bloch functions are seldom  known explicitly ...". Thus, no
explicit calculation of  matrix elements $\langle\psi_{\rm f}
|H_{\rm int}|\psi_{\rm i}\rangle$ involving subbands is known.
The explicit matrix elements are replaced by the so-called oscillator strengths $f_{\rm
fi}$, statistical assumptions and sum rules.

When the calculation of optical matrix-elements
is pursued in the standard approaches,\cite{ChangSchulman,LuoFurdyna1990,Helm1991} the calculation  ends up
evaluating at the center of the Brillouin zone or at the subband
edge,\cite{ChangSchulman,Dignam} $q=0$. With continuous subaband structures obtained from Kronig-Penney theory, and the single-quantum-well
eigenfunctions as envelope function with the ensuing
selection rule $\Delta \mu$=$\mu$-$\mu'$=0, widely used to
explain the optical transitions
structure.\cite{Dingle1975ASSP,Helm1991} Luo et
al.\cite{LuoFurdyna1990} suggested also  $\Delta \mu=0,\pm 1$ for
narrow gap materials. The limits of this approach became apparent
when optical transitions, ``forbidden"  by these rules,  were
observed,\cite{SandersChang,
Masselink,Molenkamp1988,Reynolds1988,Fu,Zhu1995} and when important features of  photoluminescence spectra, obtained in high-resolution measurements, could not be explained.\cite{NakamuraBook,Nakamura2,Nawakami1997,Nawakami1997} It is worth
noticing that, independent  of the models' limitations  to
solve the  Schr\"odinger equation for periodic potentials, the
effective-mass\cite{Slater,James,Einevoll1988}  and the
envelope-function\cite{Bastard1981,WhiteSham,AltarelliLesHuches}
approximations are important advances in the aim to work
out the quantum problem of  heterostructures and
SLs.\cite{Bastard,AltarelliLesHuches,Mailhiot,ChangSchulman,LuoFurdyna1990,Molenkamp1988,
Helm93,Virgilio}

On the other side, the theory of finite periodic systems (TFPS) grew up along the last 50 years, in the spirit of the Landauer's scattering approach,\cite{Landauer1970} and the electromagnetic theory
of periodic media.\cite{Abeles,WolfBorn} Transport properties  of
periodic semiconductor systems, modelled as  Kr\"onig-Penney
like\cite{KronigPenney}  square barriers and wells  in  the effective mass
approximation, have been studied.\cite{TsuEsaki1973,Erdos,Claro1982,Ricco,Vezzetti,RPA,Lee,Cruz1990,Kolatas,Griffiths,
Sprung,Rozman,Peisakovich,Yeh,PereyraPRL,
PereyraCM,Pereyra2000,PereyraCastillo,Assaoui,Kunold,Pacher2003,Pereyra2005,Simanjuntak}
This approach evolved and the TFPS has been generalized to include periodic structures with arbitrary potential profiles, arbitrary number $n$ of unit cells
and arbitrary number $\mathcal{N}$ of propagating modes for open, bounded and
quasi-bounded SLs.\cite{PereyraPRL,PereyraCM,Pereyra2000,PereyraCastillo,Assaoui,Pereyra2005}  As was amply explained
in these references,  important physical quantities, like the eigenvalues, eigenfunctions and transmission coefficients, are
straightforwardly obtained without recurring to Bloch's theorem, but
using simple algebraic procedures and general transfer-matrix
properties, i.e.,  with  the same elementary mathematics
that we use to solve  quantum-mechanics textbook examples.\cite{PereyraFQP}
The theory, is based on the transfer matrix method, and the important combination property that makes possible to express the n-cells transfer matrix $M_n$ as $M^n$, where $M$ is the single-cell transfer matrix. This relation, $M_n=M^n$, has been rigorously transformed into the non-commutative recurrence relation
\begin{equation}\label{RecRelation}
p_n-(\beta^{-1} \alpha \beta+ \alpha^*)p_{n-1}+p_{n-2}=0.
\end{equation}
and solved.\cite{PereyraPRL} The matrix polynomials $p_n$ of dimension $\mathcal{N}$$\times$$\mathcal{N}$ become, in the one-propagating mode approximation, the Chebyshev polynomials of the second kind $U_n$. Given these polynomials, the transfer-matrix blocks $M_n(i,j)$, for time reversal invariant systems, are: $M_n(1,1)=\alpha_n= p_n-\alpha* p_{n-1}=M_n^*(2,2)$ and $M_n(1,2)=\beta_n=\beta p_{n-1}=M_n^*(2,1)=$, with  $\alpha$ and $\beta$ the single cell transfer matrix elements. The matrix $M_n$ and the polynomials $p_n$ carry, like the scattering matrix, the whole information of the physical processes in the system, and depend explicitly on the system size, i.e. on $n$. Analytical expressions for important physical quantities, like the scattering amplitudes, $t_n$ and $r_n$, the eigenfunctions $\phi_{\mu
\nu}(z)$ and the eigenvalues $E_{\mu \nu}$, have been obtained.

It is worth to stress here that, whereas in the
infinite-periodic approaches to SLs,
\cite{Bastard,Mukherji1975,Dingle1975ASSP,SaiHalasz1978,LuoFurdyna1990,
ChangSchulman,Yang,HaugKoch,TBM}
each subband (with an infinite number of energy levels, see figure 1) is
described by a function $\chi_{\mu}(z)$ characterized by  a single
index $\mu$,  in the theory of finite periodic systems the
subbands (for a system with an arbitrary number of unit cells $n$)
are completely resolved and each subband $\mu$ is characterized by
a finite set of explicitly determined eigenfunctions $\phi_{\mu
\nu}(z)$ and eigenvalues $E_{\mu \nu}$, where the index $\nu$
labels the intra-subband levels with values $\nu$=1, 2, ...,
$n$+1.\cite{Pereyra2005,PereyraPRL} Each eigenfunction  $\phi_{\mu
\nu}(z)$, as shown in Ref. [\onlinecite{Pereyra1}], has a well defined parity
determined by the subband index  $\mu$, the intra-subband index
$\nu$ and the number of unit cells $n$.  It is clear, because of
this difference,  that not only the transition matrix elements but
also the selection rules will not coincide.

Since the TFPS is, clearly, the appropriate approach to study superlattices, and the theoretical descriptions of the optical response has been surpass by the experimental developments, we present here a detailed  discussion and explicit calculations of optical responses using  this approach. We will discuss the optical-response-calculation problem in general and  we will also apply to specific systems.

Before we focus on the optical-response calculation problem, let us highlight some issues and results that will be faced in this approach. From experimental measurements and theoretical
calculations,\cite{Miller1976,Helm1991,Nakamura2,Nawakami1997,
NakamuraBook,Narukawa1999,Kunold,PereyraAvila}  that the
subband-separations and subband-widths in SLs are of the order of
one tenth of the band offsets, with intra-subband level
separations of the order of 1 meV. Real transitions occur between
discrete states in the subbands. We will see that given a SL, one can perfectly
distinguish energy eigenvalues that differ by 10$^{-10}$eV or
less. This will allows us to describe high accuracy photoluminescence
(PL) spectra,  with peak separations of the order of 1 meV or
less, that were  identified as longitudinal
modes.\cite{Dingle1975, NakamuraBook}

An important problem that comes out when the subband structure is resolved, is
the large number of matrix elements that one has, in
principle, to evaluate. For a SL with $n$ unit cells, $n_c$
subbands in the conduction band and $n_v$ subbands in the valence
band,  the number of energy
levels (hence of eigenfunctions), for energies below the barrier-heights, is ($n$+1)($n_c$+$n_v$) and the
number of optical transitions is $N$=($n$+1)$^2$$n_cn_v$. This
means that for a SL of lenght $L\simeq1 \mu$ and unit-cells length
$l_c\sim$50nm, like in Ref. [\onlinecite{Helm1991}],  the number
of possible transitions, for $n_c$=3 and $n_v$=4, is
$N$$\sim$5,000. This is a large number. One of the purposes of this
paper is to reduce substantially this number based on new selection rules based on important symmetries.  We will show that using the eigenfunctions symmetries, derived in
Ref. [\onlinecite{Pereyra1}], we will establish the symmetry selection rules that will reduce the number of matrix-elements evaluations to $N/2$. We will then discuss a couple of rules. The leading order selection rule determined by the subbands symmetry and the edge states rules, which
reduce substantially the number of optical transitions  into one
of the order of $n_cn_v$ $\sim$10, giving essentially the same
photoluminescence (PL) and infrared (IR) results. We will apply these rules for a number of specific
examples. 

In the second section, we will outline the theoretical model for optical response of SLs
and the eigenfunctions parity symmetry relations found in Ref. [\onlinecite{Pereyra1}]. In the third section, after
writing the selection rules based on the eigenfunction parity
symmetries, we will consider three illustrative examples. We choose as
the first examples two of the Nakamura's high precision PL spectra
for the blue emitting $GaN\backslash(In_{x}Ga_{1-x}N\backslash
In_{y}Ga_{1-y}N)^{n}\backslash Al_{0.2}Ga_{0.8}N$ SLs, for $n$=10
and for $n$=7. We will show that the observed resonant spectrum and group structure, that
couldn't be explained before, will be fully explained. We will show in the
second case the effect of the cladding-layers asymmetry.
We will then consider the IR spectra of the
$(Al_{0.3}Ga_{0.7}As\backslash GaAs)^{n}$ SLs studied by Helm et
al., where the number of unit cells $n$ is of the order of 400. We will discuss the effect
of $n$ and show that equivalent results can be obtained when the
number of unit cells is, say, of the order of 20.  We conclude
section \ref{SelRules}  with a brief discussion on the characteristic resonance
line-shapes, as well as the exciton binding energies effects on
the PL spectra. As a specific example, we will consider one of the
various results reported by Masselink et al. for
$(Al_{0.3}Ga_{0.7}As\backslash GaAs)^{n}$ SLs. 

In section \ref{LOR}, we discuss the leading order rules. We will conclude the paper with a brief discussion, in section, \ref{SESR} on the surface and
edge-states rules (SESR),  implying a minimum of matrix-elements
evaluations, of the order of $n_cn_v$, with essentially the same results as with $N/2$ matrix evaluations. In the appendix we will extend the Leavitt-Little model to
include the exciton binding energy in the first excited state.

\section{The SLs optical response in the TFPS}

To study the optical transitions  in a superlattice,  in the presence of an electromagnetic (EM) field, we consider the Hamiltonian
\begin{equation}
H=H_o+H_{\small EM}+H_I,
\end{equation}
where\cite{EMA}
\begin{eqnarray}
H_o &=&-\frac{\hbar^2}{2m_e^*}\nabla_e^2 + V_e(z_e)+\frac{\hbar^2}{2m_h^*}\nabla_h^2 + V_h(z_h)  \nonumber \\ && +V_{eh}(|{\bf r}_e-{\bf r}_h|)  \hspace{0.3in} z_L \leq z_e, z_h \leq z_R
\end{eqnarray}
describes  an ({\it electron-hole}) pair in the finite periodic potentials $V_e(z_e)$ and  $V_h(z_h)$, being  $V_{eh}$ the Coulomb interaction potential. $H_{\small
EM}$ describes the transverse EM  field and $H_I$  the exciton-field interaction. In this approach, the main part of the quantum problem is centered in solving  the electron and hole SL Schr\"odinger equations
\begin{equation}\label{elSLSCHEq}
\left( -\frac{\hbar^2}{2m_e^*}\nabla_e^2 + V_e(z_e) \right) \varphi_{\mu,\nu}(z_e)=E_{\mu,\nu}\varphi_{\mu,\nu}(z_e),
\end{equation}
and
\begin{equation}\label{hoSLSCHEq}
\left( -\frac{\hbar^2}{2m_h^*}\nabla_h^2 + V_h(z_h) \right) \varphi_{\mu',\nu'}(z_h)=E_{\mu',\nu'}\varphi_{\mu',\nu'}(z_h).
\end{equation}
The wave functions for the Hamiltonian $H_o$ can be written as
\begin{equation}
\Psi_i({\bf r}_e,{\bf r}_h)=\varphi_{\mu,\nu}(z_e)\varphi_{\mu',\nu'}(z_h)\phi_{j}({\boldsymbol \rho}_e,{\boldsymbol \rho}_h,z_e-z_h),
\end{equation}
with  $\phi_{j}({\boldsymbol \rho}_e,{\boldsymbol \rho}_h,z_e-z_h)$ an eigenfunction of the quasi-two-dimensional Schr\"odinger equation
\begin{eqnarray}
\Bigl(\!-\!\frac{\!\hbar^2}{\!2m_{\parallel e}^*}\nabla_{\parallel e}^2\!\!\!&-{\displaystyle \frac{\hbar^2}{ 2m_{\parallel h}^*}}\nabla_{\parallel h}^2 \!\!-\!\! V_{eh}(|{\bf r}_e\!-\!{\bf r}_h|) \Bigr) \phi_j({\boldsymbol \rho}_e,{\boldsymbol \rho}_h,z_e\!-\!z_h) \cr &=-E^{(2D)}_j\phi_{j}({\boldsymbol \rho}_e,{\boldsymbol \rho}_h,z_e-z_h).\!\!\!
\end{eqnarray}
where $\nabla_{\parallel i}$ is the in-plane component of the gradient with respect to the two-dimensional vector ${\boldsymbol \rho}_i$. A great deal of effort has been devoted in solving this equation, which provides the {\it e-h} pair binding energies.\cite{Greene1984,Chomete1987,Bastard,Fu,HaugKoch,Pereira1990,AndreaniPasquarello,
Leavitt,Lefebvre1993,Matos} Although some consensus on the order of magnitude of these energies exists, more  specific and accurate calculations for excitons in SLs are still lacking.  We will  take into account the existing results and, in the appendix, we will recall and extend the Leavitt-Little model to determine, also, the exciton binding energy in the first excited state. We shall now recall some useful results and relations related to the superlattice Shr\"odinger equations. For a simple discussion we will focus here on type I superlattices.

The SL might be open, bounded or quasi-bounded. Since most of the specific examples imply quasi-bounded SLs,  we will restrict the detailed discussion  to this kind of systems, but we will give also the selection rules in the other cases.
\begin{figure}
\begin{center}
\includegraphics [width=240pt]{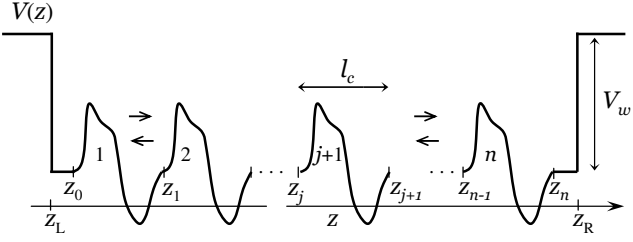}
\caption{Parameters of a quasi-bounded superlattice with arbitrary potential profile. The wave
function in Eq. (\ref{EquEigFunc2}) is defined at any point $z$ of the $j\!+\!1$
cell, with $0\leq j \leq (n-1)$.} \label{f1}
\end{center}
\end{figure}
General expressions for the evaluation of eigenvalues and
eigenfunctions of the electron and hole SL Schr\"odinger  equations (\ref{elSLSCHEq}) and (\ref{hoSLSCHEq}), were given
in Ref. [\onlinecite{Pereyra2005}]. It was shown that, provided
the single cell transfer matrix
\begin{eqnarray}
M(z_{i+1},z_{i})=\left( \begin{array}{cc} \alpha & \beta \cr \beta^* & \alpha^*  \end{array}\right),
\end{eqnarray}
is known, one can straightforwardly determine the $n$-cell transfer matrix elements, through the simple relations
\begin{eqnarray}
\alpha_n=U_n-\alpha^*U_{n-1},  \hspace{0.2in} {\rm and} \hspace{0.2in}\beta_{n}=\beta U_{n-1},
\end{eqnarray}
where  $U_n$ is the Chebyshev polynomial of the second kind and order $n$, evaluated at the real part of $\alpha$=$\alpha_R$+$i \alpha_I$. The eigenvalues of any quasi-bound SL spanning from $z_L$ to $z_R$ in figure 2, with $z_0-z_L=z_R-z_n=a/2$,  can be obtained from
\begin{eqnarray}\label{EqEigenv}
\!\!\mathfrak{R}{\rm e}\left(\alpha_ne^{ika}\right)\!-\!\frac{k^{2}\!-\!q_{w}^{2}}{2q_{w}k}\mathfrak{I}{\rm m}\left(\alpha_ne^{ika}\right)\!-\!\frac{k^{2}\!+\!q_{w}^{2}}{2q_{w}k}\beta_{nI}\!=\!0,
\end{eqnarray}
Here $q_w$ and $k$ are the wave numbers at the left (right) and right (left) of the discontinuity point $z_L$ ($z_R$),
and the eigenfunctions are given by
\begin{eqnarray}\label{EquEigFunc1}
\Psi _{\mu ,\nu }^{qb}(z)=\Psi ^{qb}(z,E_{\mu ,\nu }),
\end{eqnarray}
where
\begin{eqnarray}\label{EquEigFunc2}
\Psi^{qb}(z,E) &\!\!=\!\!&\!\frac{a_o}{2k}\Bigl[\Bigl((\alpha_{p}\!+\!\gamma
_{p})\alpha
_{j}\!+\!(\beta_{p}\!+\!\delta_{p})\beta_{j}^{\ast}\Bigr)e^{\!i k a/2}(k\!-\!iq_{w}
)\Bigr. \nonumber \\ &\!\!+\!\!&\Bigl.\Bigl((\alpha
_{p}\!+\!\gamma_{p})\beta_{j}\!+\!(\beta_{p}\!+\!\delta _{p})\alpha _{j}^{\ast
}\Bigr) e^{\!-i k a/2}(k\!+\!iq_{w})\Bigr],\nonumber \\
\end{eqnarray}
with $a_o$ a normalization constant
and $z$ any point in the $j+1$ cell, i.e. any point between $z_j$ and $z_{j+1}$, with $0\leq j \leq (n-1)$. $\alpha_j$,
$\beta_j$,...  are the matrix elements of the transfer matrix $M_j(z_j,z_0)$ that connects the state vectors $\Phi(z_0)$ and $\Phi(z_j=z_0+j l_c)$,  at points separated by exactly $j$ unit cells, and $\alpha_p$,
$\beta_p$,...the matrix elements of the transfer matrix $M_p(z,z_j)$ that connects the state vectors $\Phi(z_j)$ and $\Phi(z)$, for $z_j\leq z \leq z_{j+1}$.
\begin{figure}
\begin{center}
\includegraphics [width=240pt]{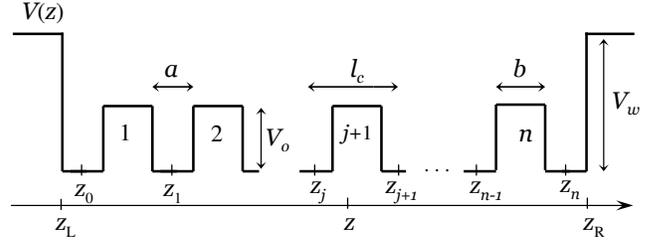}
\caption{Parameters of a quasi-bounded superlattice with with sectionally constant barriers and wells. The wave
function in Eq. (\ref{EquEigFunc2}) is defined at any point $z$ of the $j\!+\!1$
cell, with $0\leq j \leq (n-1)$.} \label{f2}
\end{center}
\end{figure}
In the particular case of a quasi-bounded periodic potential like the one shown in figure 3, the eigenvalues equation can be written as
\begin{equation}\label{EquEigenv}
h_{w}U_{n}+f_{w}U_{n-1}=0
\end{equation}
with
\begin{eqnarray}
h_{w}&=&\frac{q_{w}^{2}-k^{2}}{2q_{w}k}\sin ka + \cos ka, \cr
f_w&=&\frac{q_w^2-k^2}{2q_wk}(\alpha_{I} \cos ka-\alpha_{R} \cos ka)-\alpha_{R}\cos ka\nonumber \\ &&-\alpha_{I} \sin ka
-\beta_{I}\frac{q_{w}^{2}+k^{2}}{2q_{w}k}.
\end{eqnarray}
and
\begin{eqnarray}
\alpha\!=\!\cosh qb\!+\!i\frac{k^2\!-\!q^2}{2kq}\sinh qb,\hspace{0.15in}
\beta\!=\!-i\frac{k^2\!+\!q^2}{2kq}\sinh qb. \hspace{0.1in}
\end{eqnarray}
The wave numbers in
the various regions (wells, barriers and cladding layers) of the
heterostructure, with the appropriate changes for electrons and
holes in the conduction and valence bands, are:
$k^2=2m^*E/\hbar^{2} $; $ q^2 = 2m^*(V_o-E)/\hbar^{2} $ and $
q_{w}^2=2m^*(V_{w}-E)/\hbar^{2} $. The potential parameters $a$,
$b$, $V_o$ and $V_w$ are shown in figure \ref{f2}. To simplify the
notation we will write just $\Psi _{\mu ,\nu }^q(z)$ for $\Psi
_{\mu ,\nu }^{qb}(z)$.

It was shown in Ref. [\onlinecite{Kunold}] that writing the exciton field as
\begin{eqnarray}
\Phi_{e-h}^r(z)&=&\sum_{\mu,\nu} \Psi_{\mu\nu}^{r,c}(z) a_{\mu\nu}+\sum_{\mu',\nu'} \Psi_{\mu'\nu'}^{r,v}(z) b_{\mu\nu}^+\nonumber \\ &=&\phi_c^r+\phi_v^{r+}
\end{eqnarray}
with $ a_{\mu\nu}$ and $ b_{\mu\nu}$ the electron and hole annihilation operators and $r$ a label to denote open (o), bounded (b) or quasi-bounded (q) SLs. The radiative interband (conduction to valence band) transition contribution of the exciton-field interaction $\langle \Phi_{e-h}^r(z)|H_I|\Phi_{e-h}^r(z)\rangle$ are given by
\begin{eqnarray}
\langle \Phi_{e-h}^r(z)|H_I|\Phi_{e-h}^r(z)\rangle_{PL}=\int dz \phi_v^{r\dagger}(z){\bf A}\frac{\partial}{\partial {\bf r}}\phi_c^r(z)
\end{eqnarray}
while the intraband (conduction to conduction band) transition contribution, of the  infrared transitions (IR), by
\begin{eqnarray}
\langle \Phi_{e-h}^r(z)|H_I|\Phi_{e-h}^r(z)\rangle_{IR}=\int dz \phi_c^{r\dagger}(z){\bf A}\frac{\partial}{\partial {\bf r}}\phi_c^r(z)
\end{eqnarray}
Therefore, the photoluminescence spectrum of a superlattice in the active zone of light emitting devices is  obtained, in the golden rule approximation, from
\begin{eqnarray}\label{susceptPL}
\chi^r_{\small PL} = \sum_{\nu,\nu'\!,\mu,\mu'}f_{eh}\frac{\displaystyle \Bigl{|}\int dz
[\Psi^{r,v}_{\mu',\nu'}(z)]^*\frac{\partial}{\partial
z}\Psi^{r,c}_{\mu,\nu}(z)\Bigr{|}^{2}}{(\hbar \omega-E_{\mu,\nu}^{c}+E_{\mu',\nu'}^{v}+E_B)^{2}+\Gamma^{2}}\hspace{0.2in}
\end{eqnarray}
\begin{eqnarray}\label{susceptPL2}
\chi^r_{\small PL} = \sum_{\nu,\nu'\!,\mu,\mu'}f_{eh}\chi^{r,PL}_{\mu'\nu',\mu,\nu}
\end{eqnarray}
with energies measured from the upper edge of the valence band. Here  $E_B$ is the exciton binding energy, $\Gamma$ the level broadening energy and $f_{eh}$ the  occupation probability, which  in terms of the quasi-equilibrium distributions $f_e$ and $f_h$, becomes\cite{HaugKoch}
\begin{eqnarray}
f_{eh}=1-f_e-f_h\propto \tanh \Bigl[ \frac{1}{2 k_B T}(\hbar \omega -E_g-\mu_{eh}) \Bigr],
\end{eqnarray}
where $k_B$ is the Boltzmann constant, $T$ the temperature and $\mu_{eh}$ the total chemical potential. At low temperatures $f_{eh}$ is just a step function.
In the same way, the infrared emissions spectrum is described by
\begin{eqnarray}
\chi^r_{IR} = \sum_{\nu,\nu'\!,\mu
\geq\mu'}f_{eh}\frac{\displaystyle \Bigl{|}\int dz
[\Psi^{r,c}_{\mu',\nu'}(z)]^*\frac{\partial}{\partial
z}\Psi^{r,c}_{\mu,\nu}(z)\Bigr{|}^{2}}{(\hbar\omega-E_{\mu,\nu}^{c}+E_{\mu',\nu'}^{c})^{2}+\Gamma^{2}}\hspace{0.2in}
\label{susceptIR}
\end{eqnarray}

\begin{eqnarray}
\chi^r_{IR}= \sum_{\nu,\nu'\!,\mu \geq\mu'}f_{eh}\chi^{r,IR}_{\mu'\nu',\mu,\nu}
\label{susceptPL2}
\end{eqnarray}

Since the wave functions $\Psi _{\mu ,\nu }^r(z)$ posses well
defined parities, as was shown in  Ref. [\onlinecite{Pereyra1}], the non vanishing
contributions imply, for PL, the parity $P[\Psi_{\mu,\nu}]$
combinations $$P[\Psi^{r,v}_{\mu^{'},\nu^{'}}]=odd \hspace{0.1in}
{\rm and} \hspace{0.1in} P[\Psi^{r,c}_{\mu,\nu}]=even, $$ or $$
P[\Psi^{r,v}_{\mu^{'},\nu^{'}}]= even \hspace{0.1in} {\rm
and}\hspace{0.1in} P[\Psi^{r,c}_{\mu,\nu}]=odd,
$$ and similar relations for IR emissions, with $c$ instead of
$v$. Using these parity combinations, we will write in the next
section a set of selection rules that will be referred to as {\it
symmetry selection rules} (SSR), to distinguish from other
empirical selection rules that will be reported in section 4,
named {\it leading order rules} (LOR) and {\it surface and edge selection
rules} (SESR), related to the intra-subband eigenfunctions' symmetries, briefly discussed in the Appendix A.

\section{Symmetry selection rules}\label{SelRules}

It was shown in Ref. [\onlinecite{Pereyra1}] that the eigenfunctions of quasi-bounded SLs fulfill the symmetry relations
\begin{eqnarray}\label{QBWFSym}
\Psi_{\mu,\nu}^q(z)\!=\!\Biggl\{\begin{array}{cc} (-1)^{\nu+1}\Psi_{\mu,\nu}^q(-z) & \text{for}\hspace{0.1in} n \hspace{0.1in}\text{odd}\cr  & \cr (-1)^{\nu+\mu}\Psi_{\mu,\nu}^q(-z) & \text{for}\hspace{0.1in} n \hspace{0.1in}\text{even}  \end{array}\Biggr..
\end{eqnarray}
This clearly leads to the following selection rules. When the number of unit cells $n$ is  even, the SSR are:
\begin{widetext}
\begin{eqnarray}\label{selectrul1}
 \int dz
\Psi^{q,v}_{\mu',\nu'}(z)\frac{\partial}{\partial z}\Psi^{q,c}_{\mu,\nu}(z)\left\{ \begin{array}{llrl}
= 0 &\hspace{0.2in}{\rm when} \hspace{0.2in} P[\mu'+\nu']&=& P[\mu+\nu]\cr
\neq 0 &\hspace{0.2in}{\rm when}\hspace{0.2in} P[\mu'+\nu']&=& P[\mu+\nu+1]\end{array}\right.
\end{eqnarray}
while for $n$ odd the SSR are
\begin{eqnarray}\label{selectrul2}
 \int dz
\Psi^{q,v}_{\mu',\nu'}(z)\frac{\partial}{\partial z}\Psi^{q,c}_{\mu,\nu}(z)\left\{ \begin{array}{llrl}
= 0 &\hspace{0.2in}{\rm when} \hspace{0.2in} P[\nu']&=& P[\nu]\cr
\neq 0 &\hspace{0.2in}{\rm when}\hspace{0.2in} P[\nu']&=& P[\nu+1]\end{array}\right.
\end{eqnarray}
\end{widetext}
Similar relations are valid for IR transitions, with the
additional restrictions $\mu \geq \mu'$ and, whenever $\mu=\mu'$,
we must also have $\nu > \nu'$. The corresponding symmetry
selection rules for open and bounded SLc are given in the appendix
1. These rules reduce effectively the number $N$ of possible
transitions by, at least, a factor 1/2, i. e. to $N/2$. This is
still a large number. In the next section we will introduce other
rules and symmetries that reduce even more the number of
matrix-elements evaluations. Before we present  other rules, let
us obtain some PL and IR spectra for specific examples using the
SSR.

\begin{figure}
\includegraphics [width=240pt]{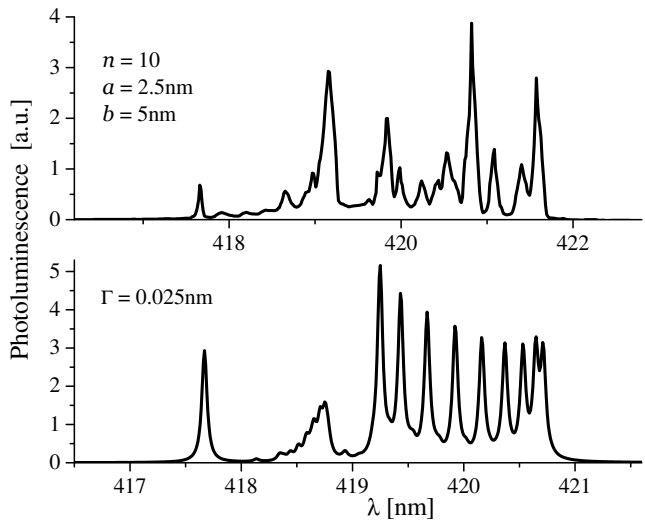}
\caption{Narrow peaks and subband groups in the PL spectra observed by Nakamura et al.\cite{Nakamura2, NakamuraBookp247} (upper panel) and calculated with this papers's technique (lower panel) for the blue emitting heterostructure $GaN\backslash(In_{0.2}Ga_{0.8}N\backslash
In_{0.05}Ga_{0.95}N)^{n}\backslash GaN$ with $n$=10, $a$=2.5nm and $b$=5nm. The SL parameters considered here are shown in figure \ref{ParSLInGaN}. The experimental spectrum is reproduced with permission
from [{\it Appl. Phys. Lett.} {\bf 68}, 3269 (1996)]. Copyright [1996], AIP Publishing LLC.}\label{Fig4}
\end{figure}
\begin{figure}
\includegraphics [width=230pt]{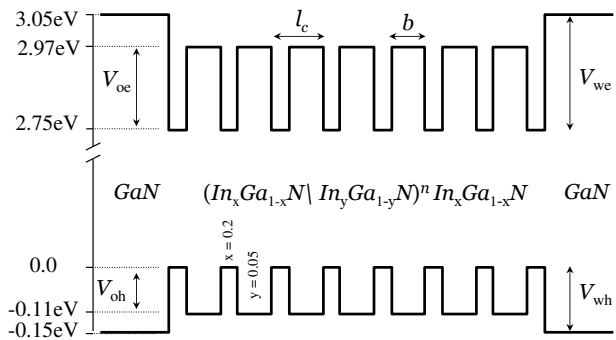}
\caption{Parameters of the $In_{x}Ga_{1\!-\!x}N\backslash
In_{y}Ga_{1\!-\!y}N$ blue emitting superlattices  used by Nakamura et al.\cite{NakamuraBook}}\label{ParSLInGaN}
\end{figure}

\begin{figure}
\includegraphics [width=210pt]{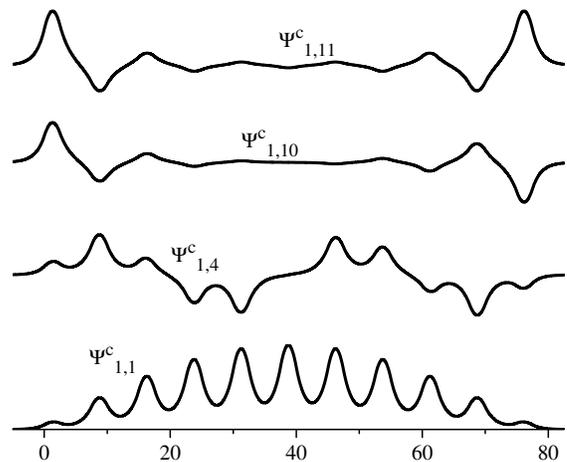}
\caption{The eigenfunctions  $\Psi _{1 ,1 }^{q,c}(z)$,  $\Psi _{1 ,4 }^{q,c}(z)$,  $\Psi _{1 ,10 }^{q,c}(z)$, and  $\Psi _{1 ,11 }^{q,c}(z)$ in the first subband of the conduction band of the blue emitting heterostructure $(In_{0.2}Ga_{0.8}N\backslash
In_{0.05}Ga_{0.95}N)^{10}\backslash In_{0.2}Ga_{0.8}N$, bounded by $GaN$ cladding layers, with  $a$=2.5nm and $b$=5nm.}\label{EigenfunctCBsb1}
\end{figure}

\subsection{PL of blue emitting devices. Nakamura's results}

As the first example we choose a  SL with the highest PL spectra
accuracy that we could find in the literature. Nakamura et al.
reported PL measurements with resolutions  of the order of
0.016nm,\cite{NakamuraBookp247} for the blue emitting
$GaN\backslash(In_{x}Ga_{1-x}N\backslash
In_{y}Ga_{1-y}N)^{n}\backslash GaN$ SLs, with $n$ varying between
3 and 20, for $x$=0.2 and $y$=0.05. In the upper panel of  figure
\ref{Fig4}, we reproduce the experimental  PL spectrum, for $n=10$. An
important characteristic of this spectrum is the presence of
narrow peaks, clustered in groups. In Nakamura's words ``it was
not clear which was the origin" of these narrow
peaks.\cite{NakamuraBookp247} Nakamura et al. suggested, that they
could originate in the ``subband transition between quantum energy
levels caused by quantum confinement of electrons and
holes".\cite{NakamuraBookp247} Our calculations, in the lower
panel of figure \ref{Fig4},  and the analysis below, show that
this is precisely the origin. The spacing and number of
narrow-peaks and  groups, correspond with the spacings of the
energy-levels and of the surface states in the subbands of the
conduction and valence bands.

\begin{figure}
\includegraphics [width=210pt]{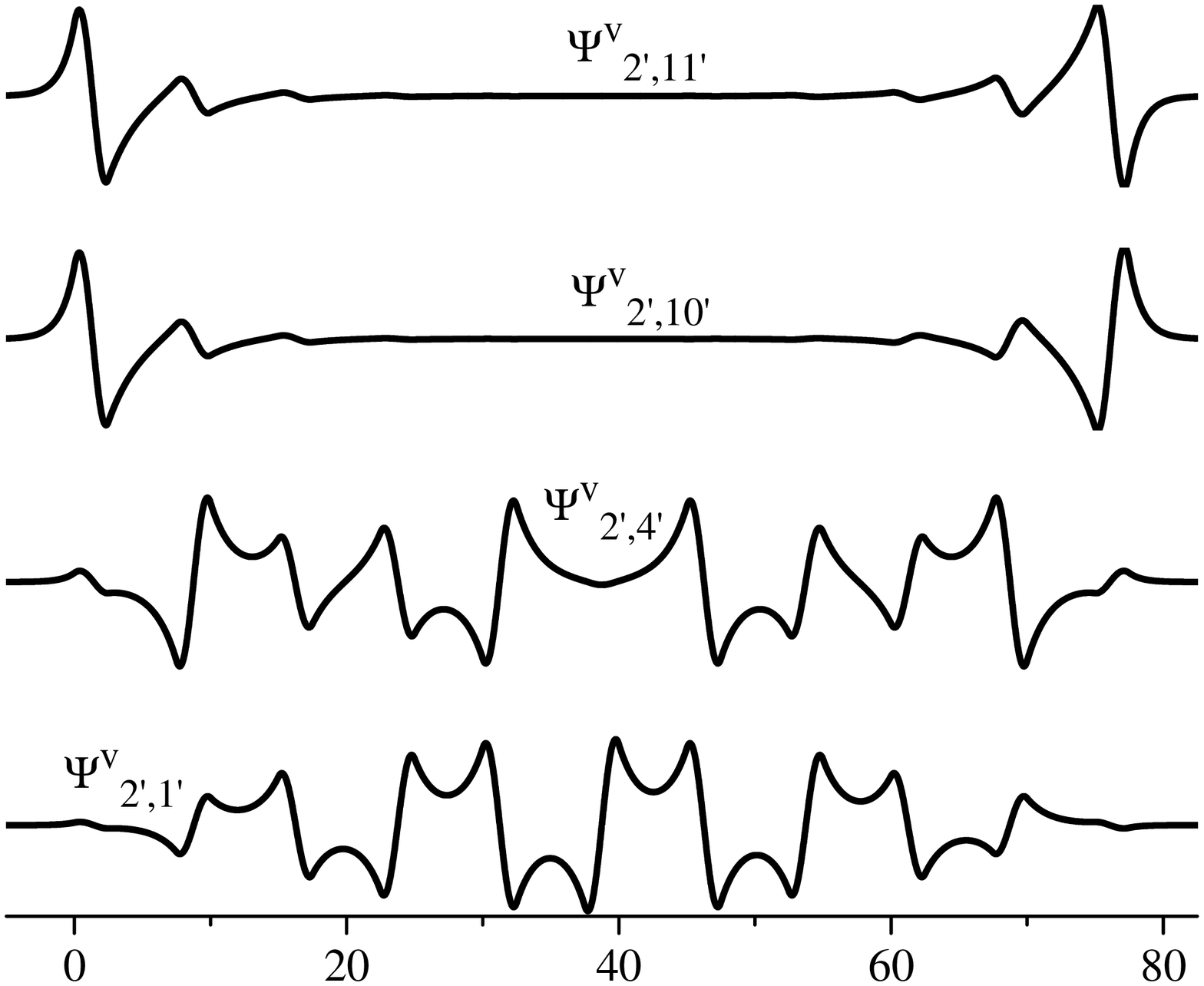}
\caption{The eigenfunctions  $\Psi _{2 ,1 }^{q,v}(z)$,  $\Psi _{2 ,4 }^{q,v}(z)$,  $\Psi _{2 ,10 }^{q,v}(z)$, and  $\Psi _{2 ,11 }^{q,v}(z)$ in the second subband of the valence band of the blue emitting heterostructure $(In_{0.2}Ga_{0.8}N\backslash
In_{0.05}Ga_{0.95}N)^{10}\backslash In_{0.2}Ga_{0.8}N$, bounded by $GaN$ cladding layers, with  $a$=2.5nm and $b$=5nm.}\label{EigenfunctVBsb2}
\end{figure}

Before we discuss our results, let us briefly sketch the steps followed for  the evaluation of  the photoluminescence spectrum for a given SL:
i) we  fix  input parameters like the well and barrier widths, gap energies, the spin split off and the effective masses;
ii) using equations (\ref{EquEigFunc1}) and (\ref{EquEigFunc2}) and (\ref{EquEigenv}), we get the eigenvalues and the eigenfunctions; iii) we normalize the eigenfunctions and evaluate the transition matrix elements, allowed by the selection rules, and iv) we plug the transition matrix elements and eigenvalues in the optical response of Eq. (\ref{susceptPL}). In Table 1 we show the energy eigenvalues $E_{1,\nu}$ in the first subband, of the conduction band and the eigenvalues $E_{2',\nu'}$ for heavy holes in the second subband of the valence band, when the SL is $(In_{x}Ga_{1-x}N\backslash
In_{y}Ga_{1-y}N)^{n}\backslash In_{x}Ga_{1-x}N$  with $x$=0.2, $y$=0.05 and $n$=10, bounded by $GaN$ cladding layers. Notice that the eigenvalues $E_{1,10}$, $E_{1,11}$, $E_{2,10}$ and $E_{2,11}$, darken in the Table are slightly detached. As was shown in Ref. [\onlinecite{Pereyra2005}], these energy levels correspond to surface states. In figures \ref{EigenfunctCBsb1} and \ref{EigenfunctVBsb2} we plot the eigenfunctions $\phi_{1,\nu}$ and $\phi_{2',\nu'}$ for $\nu,\nu'$=1, 4, 10 and 11. The last two are clearly surface states.

\setlength{\extrarowheight}{0.0cm}
\begin{center}\small
\begin{tabular}{c c c c c c}
\multicolumn{6}{c}{\it Table 1: {Energy eigenvalues $E_{1,\nu}^c$ and $E_{2',\nu'}^v$ for the SL}}  \\
\multicolumn{6}{c} { $(In_{0.2}Ga_{0.8}N\backslash
In_{0.05}Ga_{0.95}N)^{10}\backslash In_{0.2}Ga_{0.8}N$ }\\
\multicolumn{6}{c} { bounded by $GaN$ cladding layers }\\[0.05in]\hline \hline \\[-0.1in]  \;$\mu$ \;&\;$\nu $
& \!\! $E_{ \mu,\nu}^c-E_g$
\!\!  &\;$\mu'$&\;$\nu'$&$E_{\mu',\nu'}^v$ \\
\hline \\[-0.25in]
\\   &  $\;1 $  & 0.101951282609 &&\;1& -0.092052406590
\\[0.02in]    &   $\;2 $  & 0.1027359799442&&\;2& -0.092251334829
\\[0.02in]    &   $\;3 $  & 0.1040047691302&& \;3&-0.092572604852
\\[0.02in]    &   $\;4 $  & 0.1056959172240&&\;4 & -0.092999723475
\\[0.02in]    & $\;5 $  &   0.1077167669465&&\;5 & -0.093507350719
\\[0.02in]   1&  $\;6 $  & 0.1099379875843 &$2'$&\;6 &-0.094058967290
\\[0.02in]    &   $\;7 $  & 0.112187139073 &&\;7&-0.094604303220
\\[0.02in]    &   $\;8 $  & 0.1142406480441 &&\;8 &-0.095078402370
\\[0.02in]    &   $\;9 $  & 0.1158006914057&&\;9 & -0.095407111300
\\[0.02in]    &   $10 $  &{\bf 0.119074068840} &&10&{\bf -0.10012439720}
\\[0.02in]    &  $11 $  &{\bf 0.119082083156}&&11&{\bf -0.10012440024}
\\[0.02in] \hline \hline
\end{tabular}\label{Tabla10.2.1}\end{center}
\setlength{\extrarowheight}{0.1cm}

\begin{figure}
\includegraphics [width=240pt]{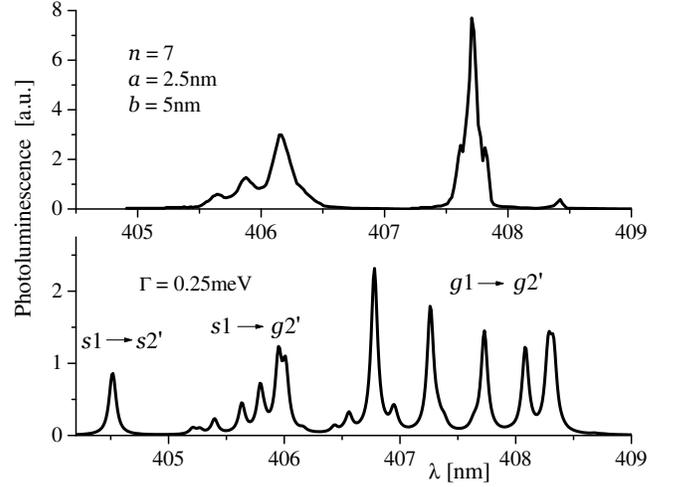}
\caption{Narrow peaks and subband groups in the PL spectra observed\cite{NakamuraBookp267} (upper panel) and calculated (lower panel) for a blue emitting heterostructure $GaN\backslash(In_{x}Ga_{1-x}N\backslash
In_{y}Ga_{1-y}N)^{n}\backslash GaN$ with $n$=7. The SL parameters  considered here are as in figure 3 but with different bowing parameters.}\label{PLExpThen7InGaN}
\end{figure}
\begin{figure}
\includegraphics [width=227pt]{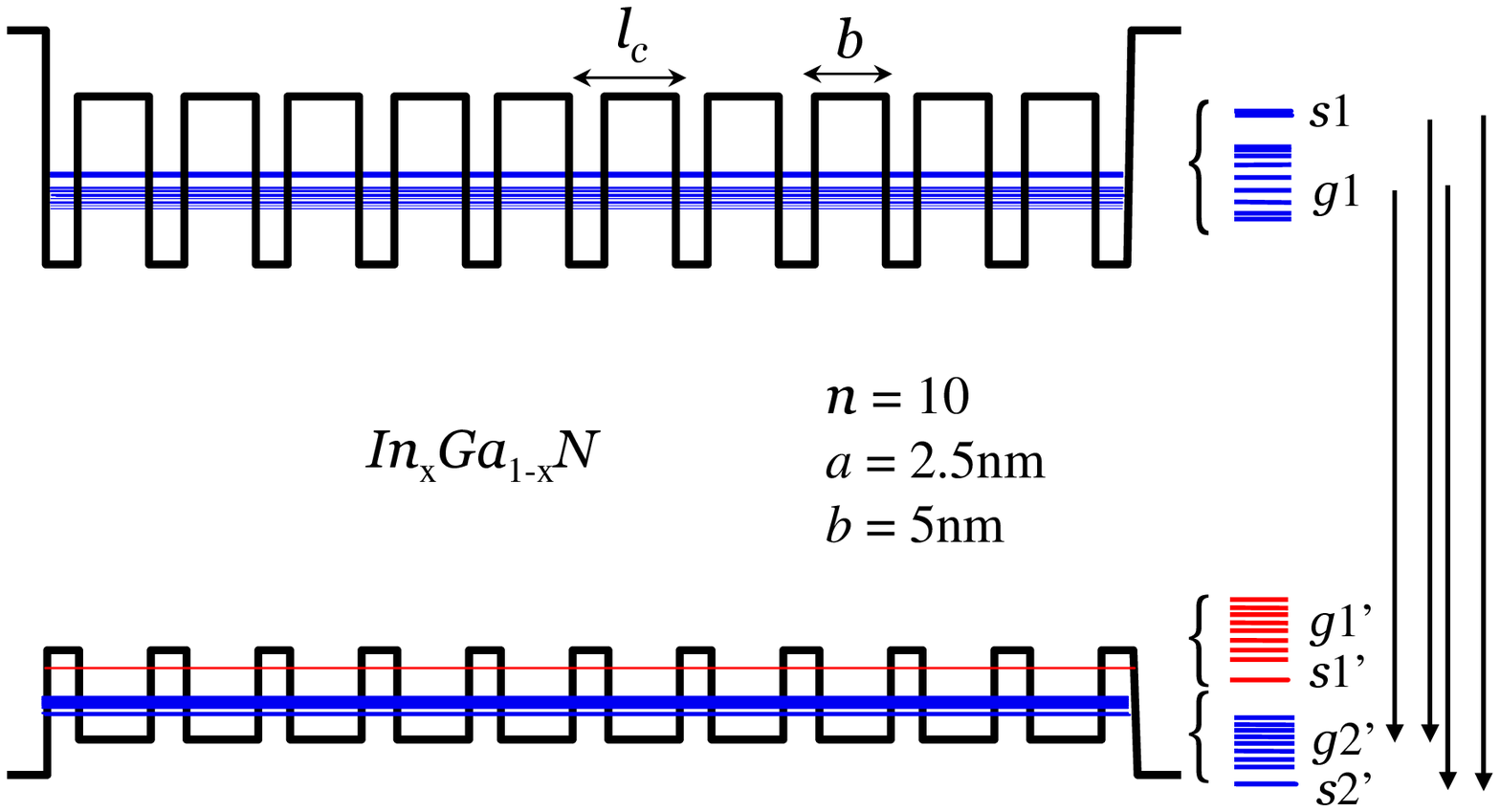}
\caption{Parameters for the blue emitting heterostructure $GaN\backslash(In_{x}Ga_{1-x}N\backslash
In_{y}Ga_{1-y}N)^{n}\backslash GaN$ .}\label{subbandSSInGaN}
\end{figure}

The experimental\cite{NakamuraBookp247} and theoretical PL spectra of this sample, with $n$=10, are shown in figure \ref{Fig4}. Similar results are shown in figure \ref{PLExpThen7InGaN} for n=7. The theoretical calculations, plotted in the lower panels, should be compared with the experimental results in the upper panels.

To understand the structure of the optical response in figures
\ref{Fig4} and \ref{PLExpThen7InGaN}, we need first to recall that
in bounded and quasi bounded system, two of the $n+1$
eigen-energies $E_{\mu,\nu}$ are surface states $SS$, which, as was
shown in Ref. [\onlinecite{Pereyra2005}] and can be seen in Table
1, detach  from the remaining $n-1$ grouped states with a
detachment energy that depends on the height of the lateral
potentials.  To simplify the reference to these states let us
denote the surface states in a subband $\mu$ as $\{s\mu\}$ and the
grouped states as $\{g\mu\}$. Similarly,  in the valence band we
have the sets $\{s\mu'\}$ and $\{g\mu'\}$, see figure
\ref{subbandSSInGaN}. For the SL with the PL spectra of figure
\ref{Fig4}, we have one subband in the conduction band and two
subbands in the valence band. The surface states in the set $s1$,
for the energy eigenvalues $E_{1,10}$=0.119074eV and
$E_{1,11}$=0.119082eV, measured from the band edge, are detached
3.2meV from the the set $g1$, grouped  in an energy interval of
13.9meV. The first subband of the valence band  is extremely thin,
the set $g1'$ is located around -0.026eV, and the SS  ${s1'}$ at
-0.028eV. This subband is so thin that, it is easy to miss it.
The set $g2'$ of the second  subband contains energy levels
$E_{2',\nu'}$ between -0.092eV and -0.0954eV, and the set ${s2'}$
is located around -0.1001244eV. The transitions sketched in figure
\ref{subbandSSInGaN} are transitions  to the second subband of the
valence band. The transitions to the first subband are negligible.
In figures  \ref{Fig4} and \ref{PLExpThen7InGaN} we have,
basically, the following transitions: $g1 \rightarrow g2'$,
responsible for the group of peaks to the right of the figure, with larger
$\lambda$s; the transitions $s1 \rightarrow g2'$, responsible for
the group in the middle and the transitions $s1  \rightarrow s2'$
responsible for the almost isolated peak at the left. The
transitions $g1 \rightarrow s2'$ are immersed in the group of
peaks in the middle.

\begin{figure}
\includegraphics [width=240pt]{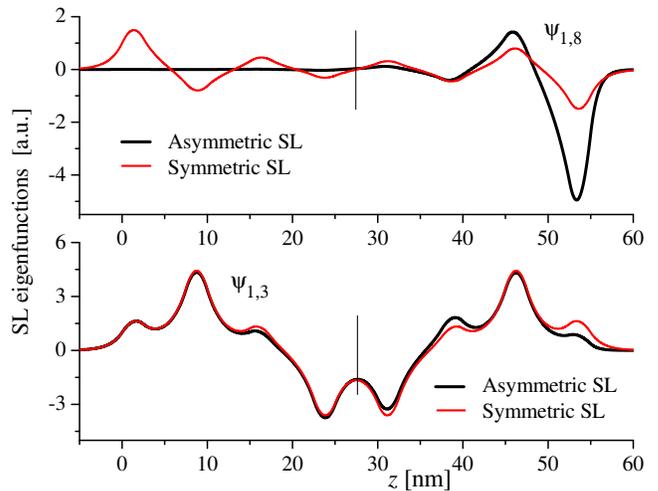}
\caption{Eigenfunctions of symmetric (red) and asymmetric (black) lateral barriers in the $(In_{0.2}Ga_{0.8}N\backslash
In_{0.05}Ga_{0.95}N)^{n}$ SL with $n$=7. The asymmetry effect is larger on the surface states.}\label{SLeigenfunctionsSAs}
\end{figure}

\begin{figure}
\includegraphics [width=240pt]{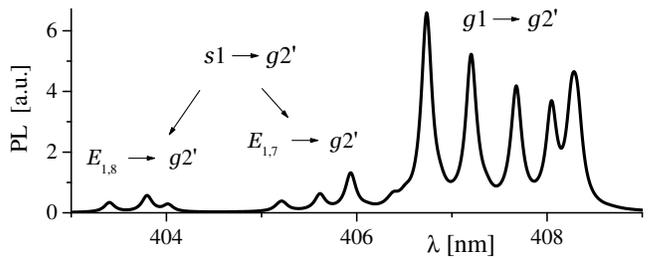}
\caption{The PL spectrum when the asymmetry of the
lateral-barriers hight is taken into account. As the surface
states are split and pushed up further,  the peaks due to
transitions $s1 \rightarrow g2'$, in the lower panel of figure
\ref{PLExpThen7InGaN}, also split, and the agreement with the
experimental PL, around $\lambda$=406nm,
improves.}\label{PLExpThen7AsymmInGaN}
\end{figure}

For this spectrum, all the matrix elements allowed by symmetry selection rules were considered. This means $N/2 \simeq 120$  matrix elements. The theoretical spectrum, based on the SL parameters shown in figure \ref{Fig4}, is slightly shifted to smaller wavelengths. A small difference like this can be fitted adjusting  parameters like the energy gaps, but this is not the goal now.  Our purpose is to stress the ability of the theoretical calculations to account for subtle features like the peaks separations and the group structure. It is worth noticing that, notwithstanding the lack of symmetry of the blue emitting SL devices, which have a $GaN$ layer on one side  and an $Al_(0.2)Ga_{0.8}N$ layer on the other,  the agreement is rather good, except for SLs with a small number of unit cells, where taking into account  the asymmetry helps.
In fact, if we consider the SL $GaN\backslash(In_{0.2}Ga_{0.8}N\backslash
In_{0.05}Ga_{0.95}N)^{n}\backslash Al_{0.2}Ga_{0.8}N$ with $n=7$, and compare with the symmetric one, we have,  related with the eigenfunctions characteristics, that while the low lying eigenfunctions of  the asymmetric and symmetric SLs (shown in the lower panel of figure \ref{SLeigenfunctionsSAs}) almost coincide, the high energy eigenfunctions (shown in the upper panel) differ. In general the asymmetry effect is larger on the surface states rather than in the low lying ones; the surface energy levels detach further and the particles get also localized. The PL spectrum of the asymmetric SL $GaN\backslash(In_{0.2}Ga_{0.8}N\backslash
In_{0.05}Ga_{0.95}N)^{7}\backslash Al_{0.2}Ga_{0.8}N$, is shown in figure \ref{PLExpThen7AsymmInGaN}. The structure reflects the fact that one of the two, almost degenerate, surface states in $\{s1\}$ is pushed up, about 15meV,  while the other states remain practically in the same position. Because of this splitting in $s_1$, instead of the transitions $s1 \rightarrow g2'$, we have now two groups: the transitions  $E_{1,7} \rightarrow g2'$ and the transitions $E_{1,8} \rightarrow g2'$,  each with three peaks. At the same time the large peak, due to the  $s1  \rightarrow s2'$ transition, practically disappears, as can also be seen in the experimental spectrum.


It is worth noticing, however, that even though the range of predicted  wavelengths for the PL peaks is of the order of the observed ones, using either the gap energy $E_{g,GaN}=3.4$ or $E_{g,GaN}=3.2$, with the appropriate bowing parameter in $E_g(x)=x1.95 +(1-x)E_{g,GaN}-Bx(1-x)$, a better agreement is found when the energy gap of the cubic (zincblende) structure, $E_{g,GaN}=3.2$, is taken into account.\cite{WhyNotChargePolariz}

An important feature in the PL spectra, particularly in the low resolution measurements, is the relatively small number of peaks. This leads us to recognize other rules that  will be introduced  in the next sections, which pick up  matrix elements that contribute more to the PL and the IR spectra.

\subsection{Infrared transitions}

Let us now consider IR spectra for another kind of SLs where the surface energy levels do not detach from the other subbands or minibands. This happens when the semiconductor in the cladding layers is the same as that of the barrier. Helm et al.\cite{Helm93,Helm94,Helm95} produced $(Al_{0.3}Ga_{0.7}As\backslash
GaAs)^{n}\backslash Al_{0.3}Ga_{0.7}As$ SLs with a large number of unit cells ($n$=200, 400 and 500) and valley and barrier widths of a few nanometers up to 40nm. They measured the IR spectra and based on these results together with  the golden rule plus the dispersion relations (calculated from the Kronig-Penney model), they were able to infer the density of states in the minibands as well as the oscillator strengths. We will now show that being able to explicitly evaluate the transition matrix elements,  the experimental spectra can be accounted for without experimental parameters and without assumptions on oscillator strengths.

\begin{figure}
\includegraphics [width=230pt]{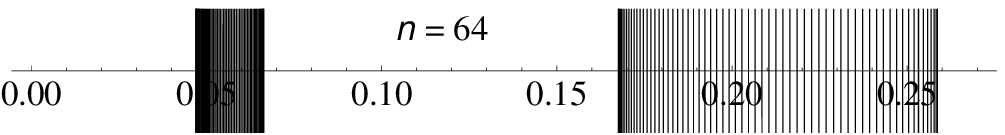}
\vspace{0.1in}
\vspace{0.1in}
\includegraphics [width=230pt]{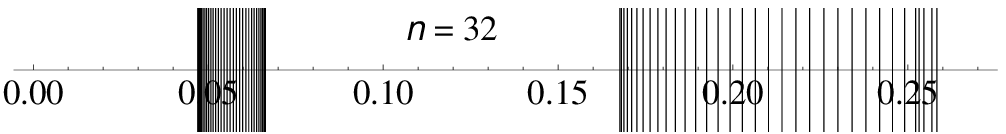}
\vspace{0.1in}
\vspace{0.1in}
\includegraphics [width=230pt]{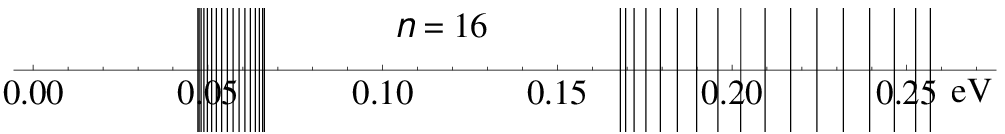}
\vspace{0.1in}
\caption{Subbands of energy eigenvalues in the conduction band of the $(Al_{0.3}Ga_{0.7}As\backslash
GaAs)^{n}$ SL. The subband widths are independent of the number of unit cells $n$; as $n$ grows, the energy-eigenvalues densities become essentially  the same up to a constant factor.}\label{subbands}
\end{figure}

Before we present our results it is worth to recall and to stress
some important properties related with the subband widths and the
density of states. It is well known that  the band widths, determined
in the TFPS by the trace of the single cell transfer matrix, i.e.
by Tr$M/2\leq1$, do not change when the number of unit cells $n$
varies from, say, $n$=20 to $n$=30 or to $n$=500. This can be seen
in figure \ref{subbands} where the energy spectra of
$(Al_{0.3}Ga_{0.7}As\backslash GaAs)^{n}\backslash
Al_{0.3}Ga_{0.7}As$ is plotted for  $n$=16, 32 and 64. It was
shown  in Ref. [\onlinecite{PereyraCastillo}] that  in the large
$n$ limit, the density of  energy eigenvalues agrees with the
density of states in the continuum derived by Kronig and Penney.
It is then clear that, once the energy eigenvalues are obtained,
their spacings define implicitly the actual density of states, and
the matrix elements (of the electron-field interaction) define, in
principle, the so called oscillator strengths.  We will show below
that the IR spectrum of a SL with $n$=20 is the same as that for
$n$=30, and describes quite well the experimental spectrum of a SL
with $n$=200. Therefore, producing SLs with 200 or 500 periods,
will no longer be necessary, though it helped to justify the use
of a theory for infinite periodic systems.

For the calculation of the IR spectrum based on the golden rule (\ref{susceptIR}) we use again the symmetry selection rules (\ref{selectrul1}) and (\ref{selectrul2}), with $\Psi^{q,v}_{\mu',\nu'}(z)$ replaced by $\Psi^{q,c}_{\mu',\nu'}(z)$. As for the PL, the calculation is simple and direct when the energy eigenvalues $E_{\mu,\nu}$ and the corresponding eigenfunctions $\Psi_{\mu,\nu}$ are known.

\begin{figure}
\includegraphics [width=240pt]{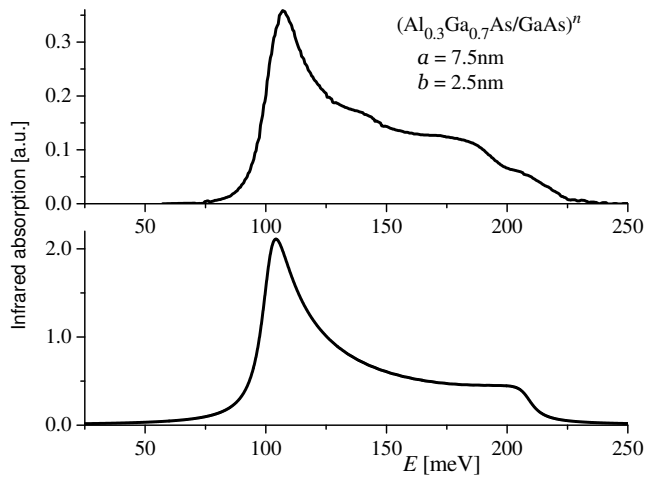}
\caption{The IR absorption spectrum measured by Helm et al.\cite{Helm93,Helm94,Helm95} (upper panel), and the theoretical calculation for the $(Al_{0.3}Ga_{0.7}As\backslash
GaAs)^{n}\backslash Al_{0.3}Ga_{0.7}As$ SL  with valley and barrier widths of 7.5nm and 2.5nm. The experimental spectrum is reproduced here with permission of the author.} \label{IRExpTheAlgAaS}
\end{figure}
\begin{figure}
\includegraphics [width=240pt]{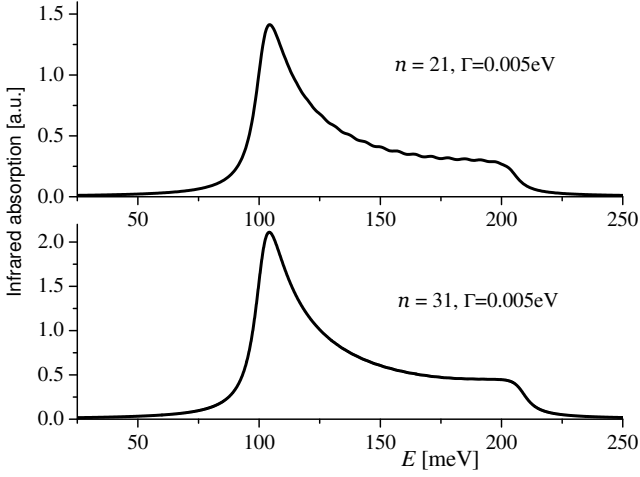}
\caption{Calculated IR spectra for the  experimental data of figure \ref{IRExpTheAlgAaS}. For these plots we considered $n$=21 (upper panel) and $n$=31 (lower panel). The results are practically the same. } \label{IRn21n31SL75w25b}
\end{figure}

In figure \ref{IRExpTheAlgAaS} we show both an experimental result
(upper panel) reported in Refs. [\onlinecite{Helm93}],
[\onlinecite{Helm94}], and [\onlinecite{Helm95}], and our
theoretical calculation (lower panel). The IR absorption was
measured for the $(Al_{0.3}Ga_{0.7}As\backslash
GaAs)^{n}\backslash Al_{0.3}Ga_{0.7}As$ SL  with valley and
barrier widths of 7.5nm and 2.5nm, respectively, and for a number
of periods of the order of 200. If we consider $n$=200, the number
of matrix elements that we need to evaluate would be of the order
of 40,000. A huge number. The experimental accuracies were,
apparently, of the order of 2-3meV and the samples were lightly
doped. For  our theoretical calculations we use the same potential
parameters as in Ref. [\onlinecite{Helm94}], and taking into
account  the arguments just explained, we considered samples with
smaller number of unit cells. Indeed,  for the IR spectra in
figure \ref{IRExpTheAlgAaS} we had $n$=31. In figure
\ref{IRn21n31SL75w25b}, we show the IR spectra for $n=21$,
implying the evaluation of 240 matrix elements. The results are
practically equivalent.   It is worth noticing that in the
theoretical calculation by Helm et al., based on the Kronig Penney
model, the first subband ranges from 37 to 55meV and the second
from 148 to 222meV. In our calculation  (see figure
\ref{subbands}),  the first subband ranges from 47 to 66meV and
the second from 168 to 254meV. This implies an IR spectrum that
extends from 102meV to 207meV, which is precisely the energy
interval where the observed IR and the theoretical spectrum for
$\Gamma$=0.005eV lie.

\begin{figure}
\includegraphics [width=240pt]{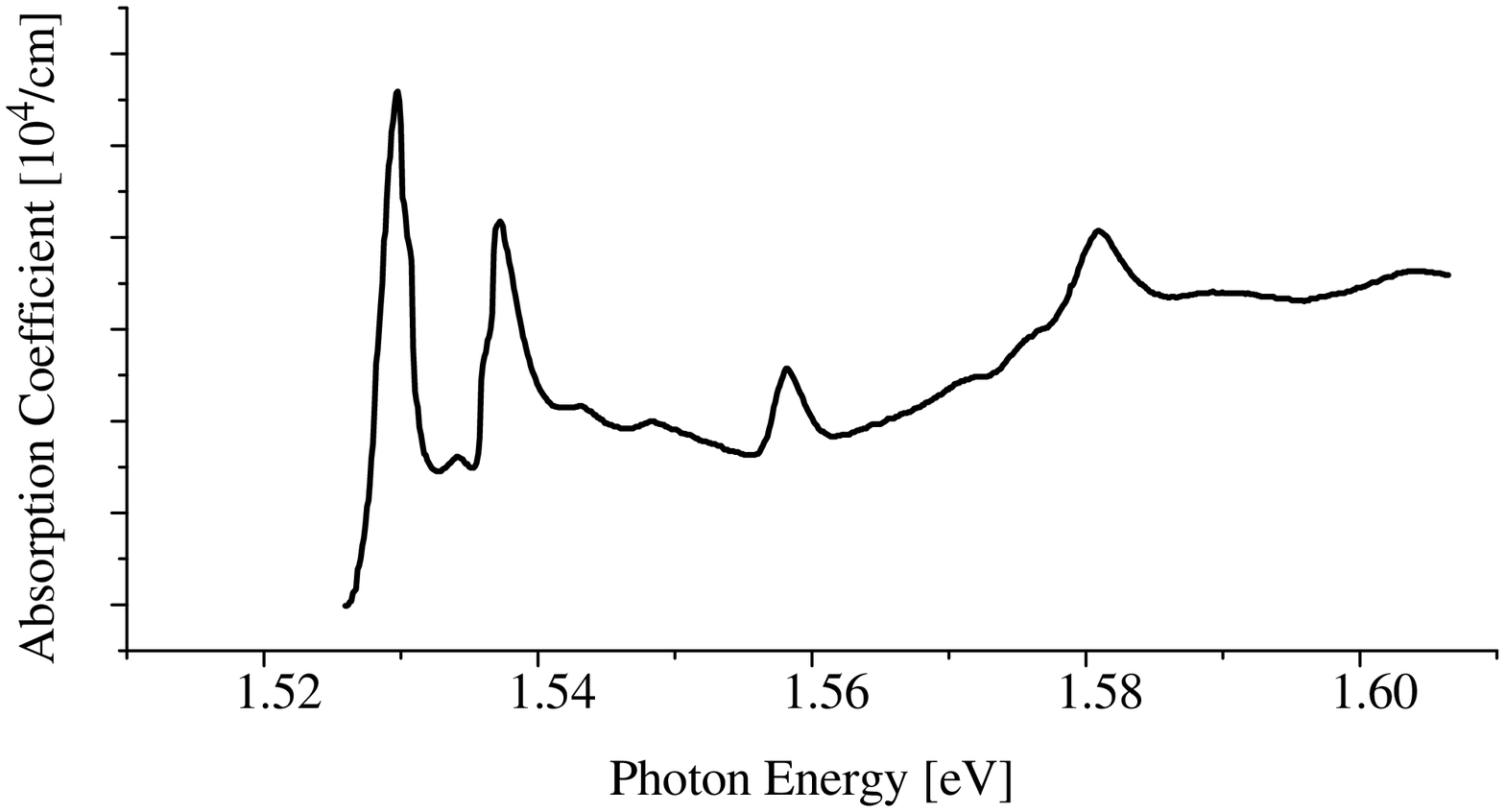}
\includegraphics [width=240pt]{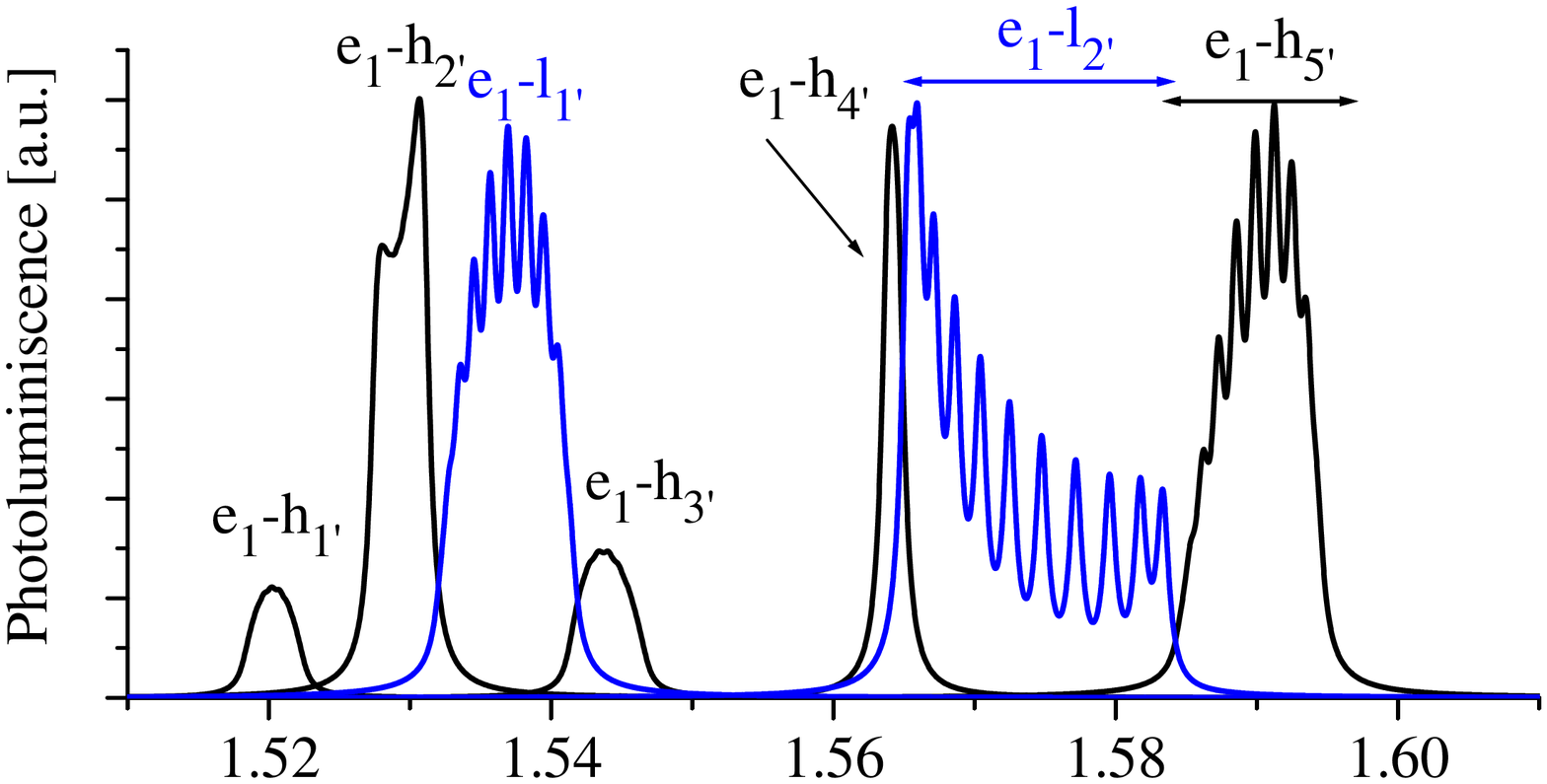}
\includegraphics [width=240pt]{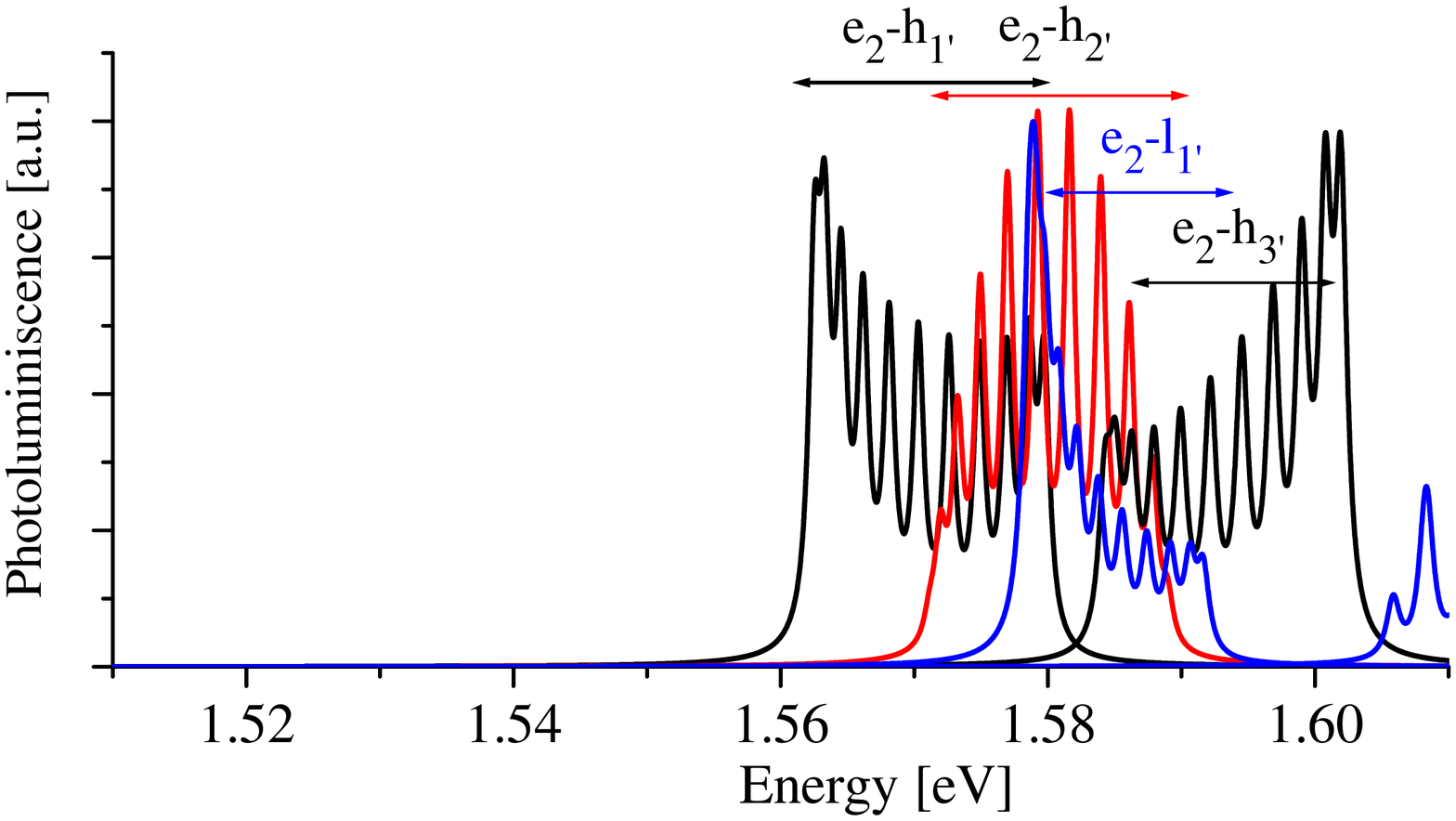}
\caption{In the upper panel absorption spectra reported by
Masselink et al. for one of the $(Al_{x}Ga_{1-x}As\backslash
GaAs)^{n}$ SLs, with well width $w$=15nm barrier width $b$=2.5nm
and SL length of the order of 1$\mu$m.  The panel in the middle
contains only the transitions $e_{1\nu}$-$hh_{\mu'\nu'}$ (black)
and $e_{1\nu}$-$lh_{\mu'\nu'}$ (blue), denoted as
$e_{1}$-$h_{\mu'}$ and  $e_{1}$-$l_{\mu'}$ (with origin in the
first subband of the CB), and in the lower panel, the transitions
$e_{2\nu}$-$hh_{\mu'\nu'}$ and $e_{2\nu}$-$lh_{\mu'\nu'}$ starting
in the second subband of the CB. For these graphs, only the ground
state exciton binding energy was considered.} \label{PLMasselink}
\end{figure}

\subsection{PL spectra for $(Al_{x}Ga_{1-x}As\backslash
GaAs)^{n}$ SLs. Exciton binding energies}

Absorption and PL spectra for $(Al_{x}Ga_{1-x}As\backslash
GaAs)^{n}$ SLs were  measured long ago by many people, among
others: Dingle et al.,\cite{Dingle1974} Miller et al.,\cite{
Miller1976} Molenkamp et al.\cite{Molenkamp1988} and Masselink et
al.\cite{Masselink} In most of the multi quantum well  structures considered in those years,  the barrier
widths were extremely large, of the order of 20 or 30nm, with
rather narrow valley widths, of the order of 5nm. The number of
unit cells were also extremely large, 100 up to 400. The purpose
was to produce a large number of independent quantum wells. We are
not going to discuss much on these type of SLs, we will instead
study below, with more detail, the PL spectrum of
$(Al_{x}Ga_{1-x}As\backslash GaAs)^{n}$ SLs with thinner barrier
widths. Wide and high barriers lead to almost independent single
quantum wells and to extremely narrow minibands, difficult to find
and to evaluate theoretically. A rigorous calculation of the
miniband structure for the sample 4-10-74 of Dingle et al., with
$x$=0.2, valley width  of 9.2nm and barrier width of 34.5nm, gives
a first miniband with $E_{1,1}$ at 0.0363318072eV, above the CB
edge, and $E_{1,n+1}$ at 0.0363318085eV. This means a subband
width of 13$\times$10$^{-7}$meV. Thus, extremely narrow minibands
on top of a wide background. In this class of systems the
theoretical calculation of surface states  is also numerically
unstable.

Masselink et al.\cite{Masselink} considered   samples with thinner barriers and wider wells. Although this, in principle, implies a larger number of subbands, with subband-widths of the order of 10meV, the theoretical calculations are feasible.   In the upper panel of figure \ref{PLMasselink} we show the spectrum reported by Masselink et al.  for the SL $(Al_{0.25}Ga_{0.75}As\backslash
GaAs)^{n}\backslash Al_{0.25}Ga_{0.75}As$  with $a$=21nm, $b$=10nm and superlattice-thickness of the order of 1 to 2$\mu$m. This means SLs with $n$ of the order of 50. In this sample, because of the wide valley width, there is a large number of subbands both in the CB and the VB. For energies below the barrier heights,  the CB contains 4 subbands; the VB contains  6 heavy-hole subdands and 3 light-hole subbands. The subband widths are, generally, smaller than 10meV. Besides the large density of subbands, there is also a high density of energy levels in the subbands. Thus, in the actual spectrum, with accuracies much larger that 0.1meV,  the single transitions peaks overlap. In the theoretical calculations there is no problem resolving the subband structures. To avoid the huge number of matrix elements and, as discussed before, without any relevant change in the results we consider $n$ of the order of 10.

In the lower panels of figure \ref{PLMasselink} we present the
theoretical calculations, black and red curves for {\it e-hh}
transitions  and blue for {\it e-lh} transitions.  To visualize
better the peaks of the possible contributions, the panel in the
middle contains only the transitions $e_{1\nu}$-$hh_{\mu'\nu'}$
(black) and $e_{1\nu}$-$lh_{\mu'\nu'}$ (blue), denoted as
$e_{1}$-$h_{\mu'}$ and  $e_{1}$-$l_{\mu'}$, with initial state in
the first subband of the CB, and in the lower panel, the
transitions $e_{2\nu}$-$hh_{\mu'\nu'}$ and
$e_{2\nu}$-$lh_{\mu'\nu'}$ starting in the second subband of the
CB. For these graphs, only the exciton binding energy in the
ground state was considered. The transitions
$e_{1\nu}$-$hh_{1'\nu'}$ and $e_{1\nu}$-$hh_{3'\nu'}$, denoted
$e_{1}$-$h_{1'}$ and $e_{1}$-$h_{3'}$ in the upper graph, have
been multiplied by large factors to make them visible.

\begin{figure}[t]
\includegraphics [width=240pt]{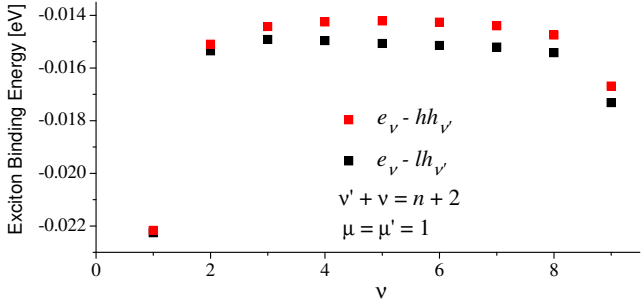}
\caption{Binding energies obtained for the superlattice $(Al_{x}Ga_{1-x}As\backslash
GaAs)^{n}$ considered in figure  \ref{PLMasselink}, after evaluating the excitons size $[\langle \mu', \nu'|(z_e-z_h)^2|\mu,\nu\rangle)]^{1/2}$ and assuming quasi-spherical excitons.}\label{BEnergQS}
\end{figure}
Before we comment our results, it is worth stressing again the
important difference between this approach and the standard one.
In the experimental reports it is rather frequent to find
statements related with the observation of ``forbidden
transitions".\cite{SandersChang,
Masselink,Molenkamp1988,Reynolds1988,Fu,Zhu1995} Forbidden in the
standard approach where only one index $\mu$ defines the ``subband
wave function" parity.  In the TFPS  there is no such thing like
the  ``subband wave function". As we have already seen, the
subbands in the TFPS contain many  intrasubband levels (all of them
with their corresponding eigenfunction characterized by two
indices) which, together with the number of unit cells, define the
selection rules. Therefore, transitions with subband indices $\mu
\neq \mu'$, forbidden in the standard theory, are perfectly
possible in this TFPS.  In our plots, the occupation probability
is assumed as a step function. No Sommerfeld effect is considered,
but the exciton binding energy was taken into account. The 2D
exciton binding energy was evaluated, as explained in the Appendix
1, from
\begin{eqnarray}\label{BindingEnerg}
E_{Bi}=-\frac{\alpha_{ri}^2 \hbar^2}{8m_r}\hspace{0.2in}{\text with}\hspace{0.2in}\alpha_{ri}=\frac{m_re^2}{2\pi \epsilon \hbar^2 \lambda_i},
\end{eqnarray}
where $m_r$ is the relative mass, $\epsilon$ the dielectric
constant and $\lambda_i$ the quantum number with values
$\lambda_0=1/2$ and $\lambda_1=3/2$, for the ground and excited 2D
exciton states, respectively. Using these energies (-18meV for
$hh$ and $-12$meV for $lh$, in the ground state) the agreement
with the experimental measurements is good. However, had we
considered binding energies of the order of
$-10$meV,\cite{SandersChang,Chomete1987,Yu1987,Reynolds1988,Dignam,Pereira1990}
 the agreement would have not been that good.

\begin{figure}
\includegraphics [width=240pt]{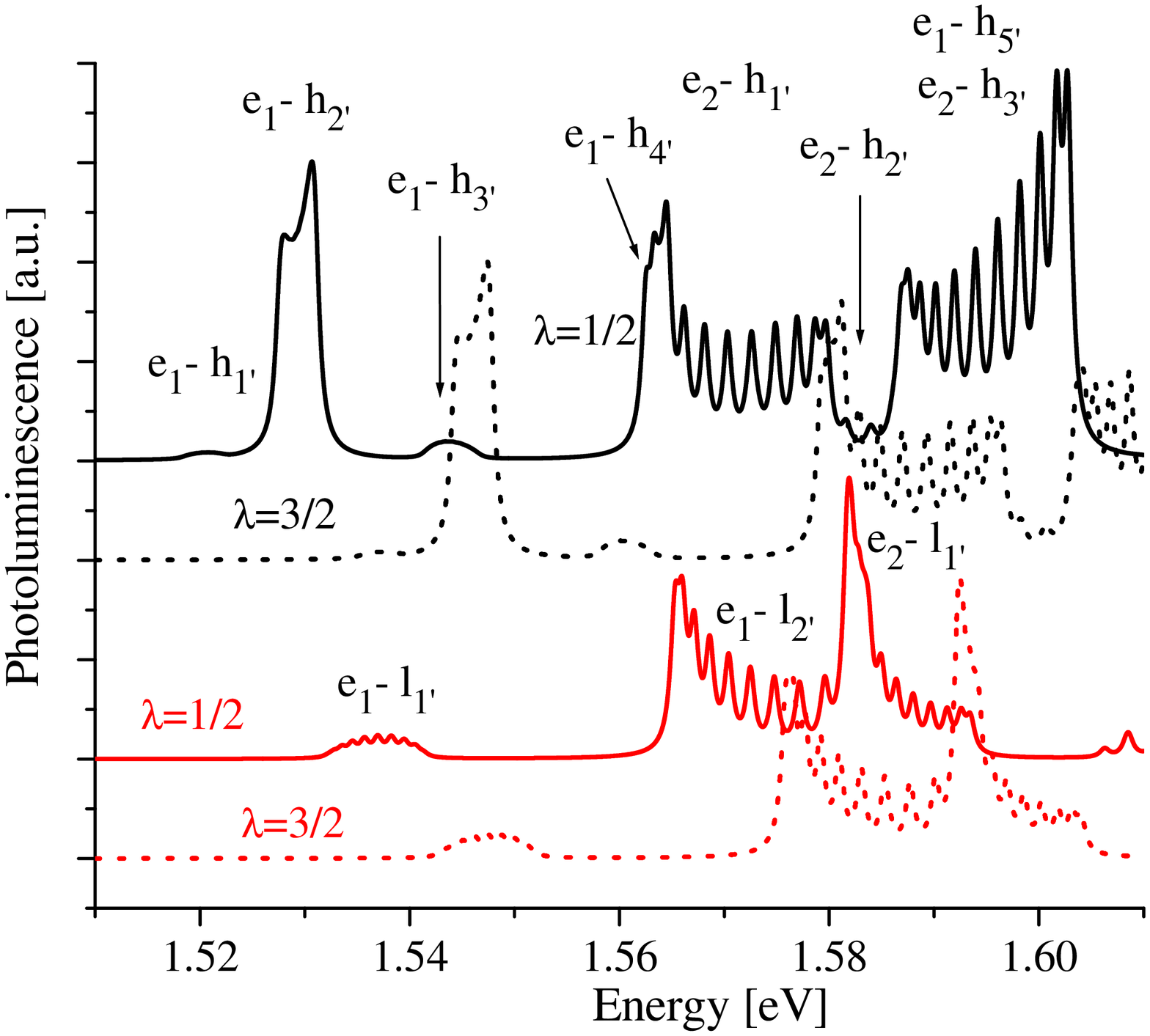}
\caption{In this figure we show again the optical transition for the superlattice $(Al_{x}Ga_{1-x}As\backslash
GaAs)^{n}$ considered in figure  \ref{PLMasselink}. Together with the 2D excitonic ground state transition (full curves) we plot also those of the first excited 2D exciton state (dashed curves). In the uppermost graphs (black curves) we have the {\it e-hh} transitions, denoted as e$_{\mu}$-h$_{\mu'}$, and in the lower graphs (red curves) we have the {\it e-lh} transitions, denoted as  e$_{\mu}$-l$_{\mu'}$. The binding energies in the 2D exciton ground state ( $\lambda$=1/2) are 0.01877meV and 0.01187meV for heavy and light holes, while in the excited state ($\lambda$=3/2) they are 0.0021meV and 0.0013meV, for heavy and light holes respectively.} \label{Masselink1}
\end{figure}
To confirm the order of magnitude of the binding energies used here, we did the following exercise. Using the SL eigenfunction, we obtain on one side the mean positions $\langle \mu', \nu'|z_e|\mu,\nu\rangle=\langle \mu', \nu'|z_h|\mu,\nu\rangle=0$, when the origin is  in the center of the SL, and, on the other side, the exciton size  along the SL from $[\langle \mu', \nu'|(z_e-z_h)^2|\mu,\nu\rangle)]^{1/2}$. Assuming quasi-spherical excitons, the binding energies take the values shown in figure \ref{BEnergQS}. Although the order of magnitude of these energies is correct, we just considered the binding energies from Eq. (\ref{BindingEnerg}).

In figure  \ref{Masselink1} we plot, besides the ground state transitions (full curves), transitions from the first excited excitonic states (dotted curves).  Within each graph of this figure, we respect the relative magnitudes except for the transitions $e_{1}$-$h_{1'}$ and  $e_{1}$-$h_{3'}$ that were again enhanced in order to visualize them. The  transitions from the excited excitonic states spectra  that we would like to notice are those corresponding to  $e_1^*$-$h_2'^*$, around 1.547eV, and $e_1^*$-$h_3'^*$, around 1.56eV,  because these may  explain those peaks in the experimental spectrum, above 1.54eV and below 1.56eV.

\begin{figure}
\includegraphics [width=240pt]{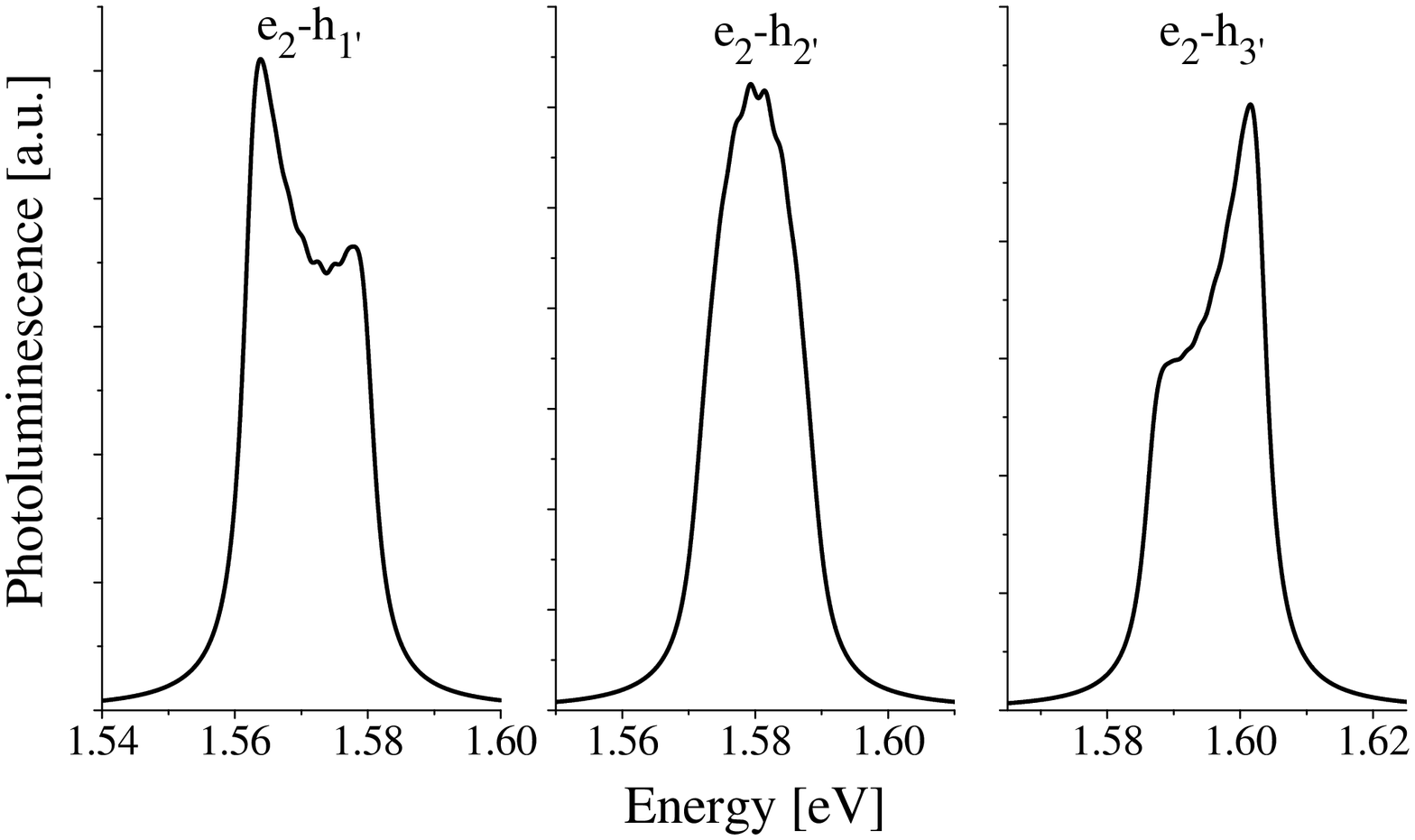}
\includegraphics [width=240pt]{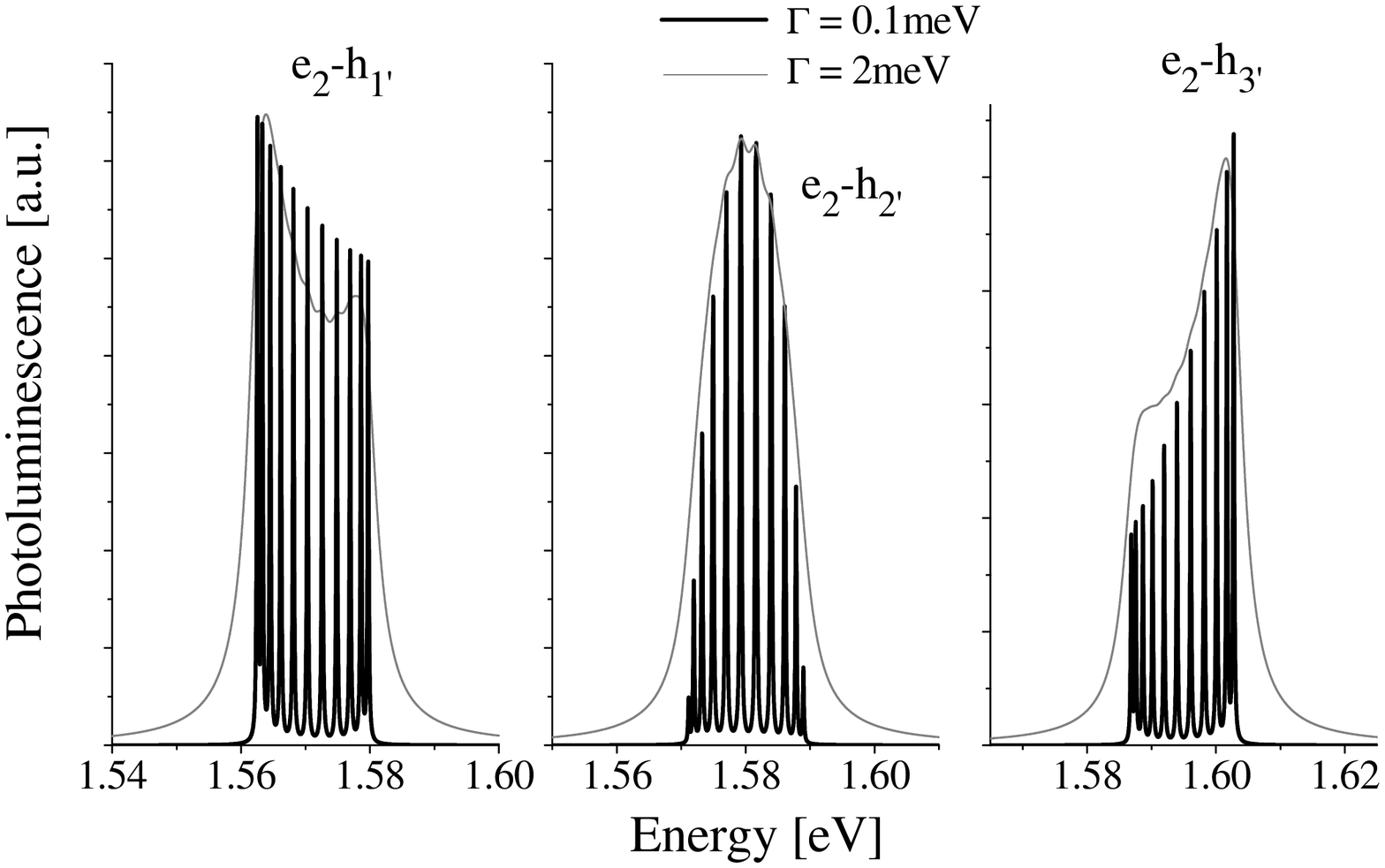}
\caption{ Lineshapes of the transitions $e_{2}$-$h_{\mu'}$ shown
in the lower frame of figure \ref{PLMasselink}, plotted here for two values of the level broadening
energy $\Gamma$. In the upper graphs we employed $\Gamma$=2meV
while in the lower graph we plot for both  $\Gamma$=0.1meV and  $\Gamma$=2meV. This makes clear that with
low accuracy experiments and calculations the intrasubband
structures disappear. These transitions, experimentally observed,
are forbidden in the standard theory.} \label{SecondMasselink}
\end{figure}
Let us now comment a couple of interesting issues. The resonances
lineshapes for transitions like $e_{\mu}$-$h_{\mu'}$ or
$e_{\mu}$-$l_{\mu'}$  have some interesting characteristics. We
found that when the subband indices $\mu$ and $\mu'$ are both even
or both odd the spectrum has a bell shape, however when they have
different parities, see figure \ref{SecondMasselink}, the spectrum
has, depending on the level broadening energy $\Gamma$, a U shape
or a  trapezoidal shape, with higher peaks at the left or right of
the spectrum. For larger $\Gamma$ or low experimental accuracy,
the spectrum  looks as having two peaks; see for example the
spectrum for $e_{1}$-$h_{2'}$. This kind of results and behavior
lead easily  to wrong assignments of subband-indices. Spectrum
shapes like these are frequent in the literature. For example, the
absorption coefficients reported by Helm et al in Refs.
[\onlinecite{Helm93},\onlinecite{Helm95}]. In these references the
two maxima are considered  critical points of the mini-Brillouin
zone and, to fit the curves, it was necessary to assume oscillator
strengths of the order of 0.3 at the center and 2.3 at the edge of
the Brillouin zone.\cite{Helm93}  In
the PL spectra of figure \ref{Masselink1}, the subband widths
$\Delta E_2^c$ and $\Delta E_3^c$,  of the second and third
subbands in the CB, are $\simeq 0.018$eV and $0.043$eV,
respectively, while the  first four subbands in the VB lie between
$\simeq 0.00293$eV and $\simeq 0.0274$eV.  Due to these
characteristics, the interband transitions $e_{2}$-$h_{1'}$,
$e_{2}$-$h_{2'}$ and $e_{2}$-$h_{3'}$ overlap. Thus, we can have,
for example, peaks of $e_{2}$-$h_{2'}$ transitions  at  higher
energies than peaks of the  transitions $e_{2}$-$h_{3'}$, see
lower panel in figure \ref{PLMasselink}. In the Kronig-Penney like
approach of Masselink et al.,  the fulfilment of boundary
conditions for even and odd solutions in the valley and barrier
led to  an "eigenvalue equation", that generalizes, slightly,  the
single quantum well eigenvalue equation.
This is not of course the correct eigenvalue equation for a
superlattice. Although the eigenvalues $E_{\mu}$ may lie inside
the subband $\mu$, one can not define with them the subband widths
$\Delta \mu$=$E_{\mu,n+1}$-$E_{\mu,1}$.

As mentioned before, a peculiar  and rather general characteristic
of most of the published and calculated PL and IR spectra, is the
small number of peaks, much smaller than the $N/2$ non-vanishing
optical transitions. One reason is of course the low experimental
precision. However, a closer look to the numerical values of the
allowed matrix elements  $\chi^{r,PL}_{\mu'\nu',\mu,\nu}=\langle
{\mu',\nu'}|\frac{\partial}{\partial z}|{\mu,\nu}\rangle$,  shows
that most of them are negligible, and only a small fraction, the
leading order transitions (LOT), determine the shape of the
optical spectrum and lead to new rules, related with the intraband
eigenfuntion's symmetry noticed in Ref.
[\onlinecite{Pereyra2005}]. This fact motivates the discussion and
content of the next section. We will repeat  the PL calculations
for the SL in the first example of this section and we will
perform PL and IR calculations for other systems as well. The
purpose is to show that, besides the SSR mentioned in this
section, there are other rules that can be introduced without
changing significantly the PL spectrum. Using the new rules we
will obtain practically the  same PL and IR spectra, with much
less effort. They can be extremely useful when the number of
subbands is large, say, two, three or more, and the number of
cells is large.

\section{The intraband symmetries. Leading order  rules}\label{LOR}

Analysis of the matrix-elements for a large number of specific
examples shows generally that the matrix elements $\langle
{\mu',\nu'}|H_I|{\mu,\nu}\rangle$ where the parity of the index
$\mu$ is the same as that of $\mu'$ are one or more orders of
magnitude smaller than the others. This means, for example, that
$$\langle {1',\nu'}|H_I|{1,\nu}\rangle << \langle
{2',\nu'}|H_I|{1,\nu}\rangle$$ or $$\langle
{3',\nu'}|H_I|{1,\nu}\rangle << \langle
{2',\nu'}|H_I|{1,\nu}\rangle .$$ On the other hand, when  the
eigenfunctions $\Psi_{\mu,\nu}$ of a given subband are plotted for
$\nu$=1, 2,..., $n$+1, we can find, as  noticed in Ref.
[\onlinecite{Pereyra2005}], another symmetry. We find that the
envelope of $\Psi_{\mu,\nu}$ is similar to that of
$\Psi_{\mu,n-\nu}$, when the surface energy levels are detached,
and similar to that of $\Psi_{\mu,n+2-\nu}$ when the surface
levels do not detach; see the appendix A. This eigenfunction symmetry related to the intraband
indices $\nu$ and $\nu'$ together with the parities of
$\mu$ and $\mu'$ define new rules that help  to determine the
matrix elements that are two or more orders of magnitude greater.
When the  potential height in the cladding layers is larger
than the barrier height in the SL, and the surface states separate
significantly,  the matrix
elements that fulfill the conditions
\begin{eqnarray}\label{LORDetached}
\langle {\mu',\nu'}|\frac{\partial}{\partial z}|{\mu,\nu}\rangle
\hspace{0.1in} {\rm with}
\hspace{0.1in} \left\{\!\!
\begin{array}{rcl}
|\mu-\mu'|\!&\!\!\!=\!\!\!&\!1,3,5,...\cr & {\rm and} &\cr
\nu+\nu'\!&\!\!\!=\!\!\!&\!n \cr \nu=n,n+1&&\nu'=1,2,... \cr
\nu'=n,n+1&&\nu=1,2,...,\end{array} \right.
\end{eqnarray}
besides the SSR, are the leading order
transitions. But when the potential height is comparable
with the barrier height in the SL, the leading order transitions correspond
to matrix elements satisfying the rules
\begin{eqnarray}\label{LORUndetached}
\langle {\mu',\nu'}|\frac{\partial}{\partial z}|{\mu,\nu}\rangle
\hspace{0.1in} {\rm where}\hspace{0.1in} \left\{
\begin{array}{rcl} |\mu-\mu'|&=&1,3,5,...\cr & {\rm with} &\cr \nu+\nu'&=&n,n+2 \end{array}
\right.
\end{eqnarray}
in addition, of course, of the SSR.   These rules that will be referred to as the leading order rules (LOR) reduce the number of evaluations for PL from $N/2\simeq(n+1)^2n_cn_v/2$ to $n n_cn_v/2$ in the first case (equation \ref{LORDetached}), and to $(n+1)n_cn_v/2$ in the last one (equation \ref{LORUndetached}). Generally the experimental results show only a part of the spectra, accordingly the number of evaluations for a given plot is also a fraction of these numbers. Therefore, in the following the actual number of evaluations behind each plot depends on how many subbands of the conduction and valence bands are taken into account.  For IR transitions we have the additional condition $\mu'\leq\mu$. The number of evaluations is reduced also significantly.\cite{reductions IR}

\begin{figure}
\includegraphics [width=240pt]{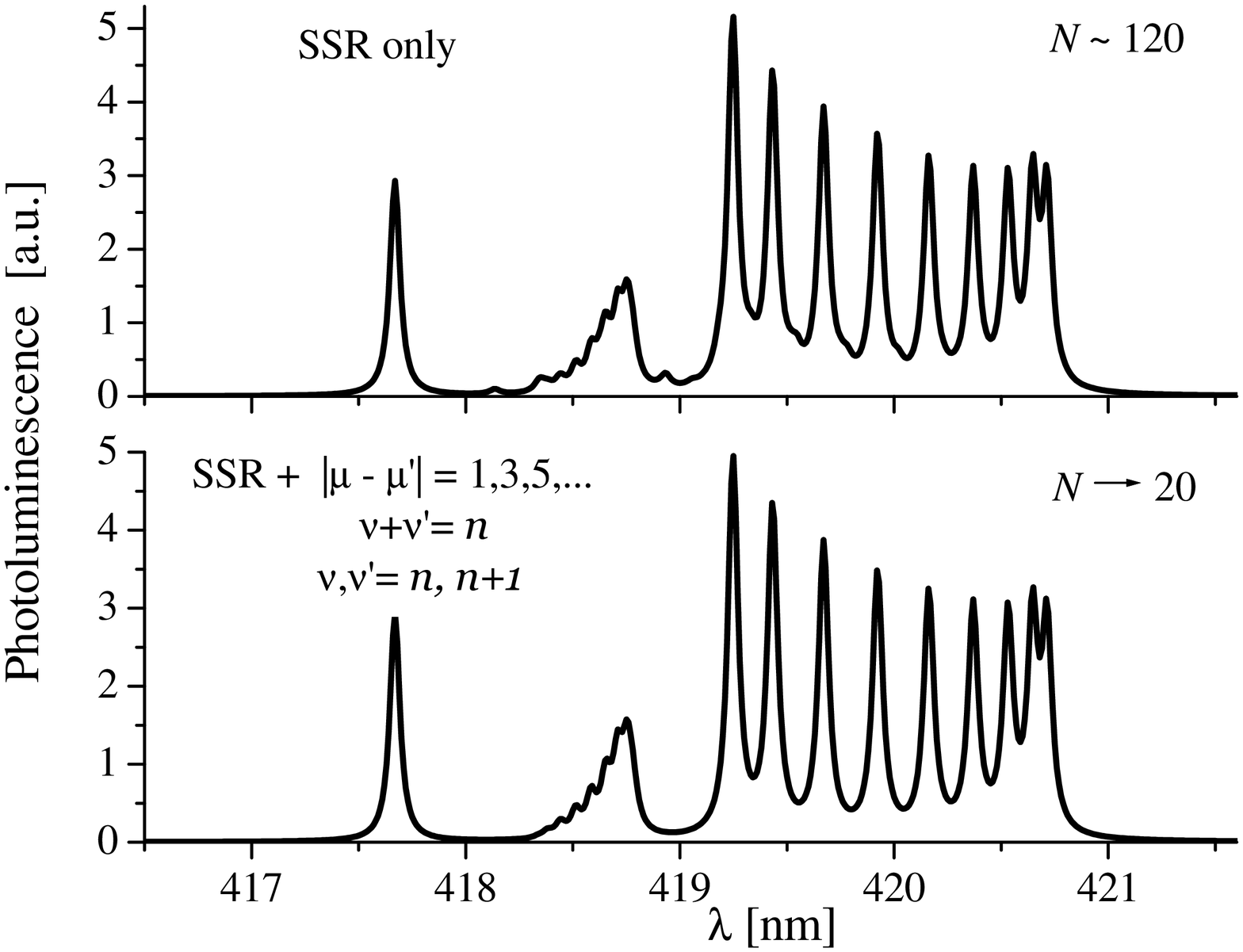}
\caption{PL spectra of the blue emitting InGaN system studied in the previous section. Here we plot the spectra with (lower) and without (upper) LOR. We show also the number of matrix elements that were calculated in each case. }\label{figure18}
\end{figure}

As an example of the first case we can consider the blue emitting system studied in the last section. For this system, with $n_c=n_v=2$ and $n=10$, the number of matrix evaluations reduces from $(n+1)^2n_v/2$=121 to $nn_vn_c/2$=20. In figure \ref{figure18}, we plot the spectra with and without the LOR. The spectrum in the lower panel contains, essentially, the same information as that in the upper one.
\begin{figure}[h]
\includegraphics [width=220pt]{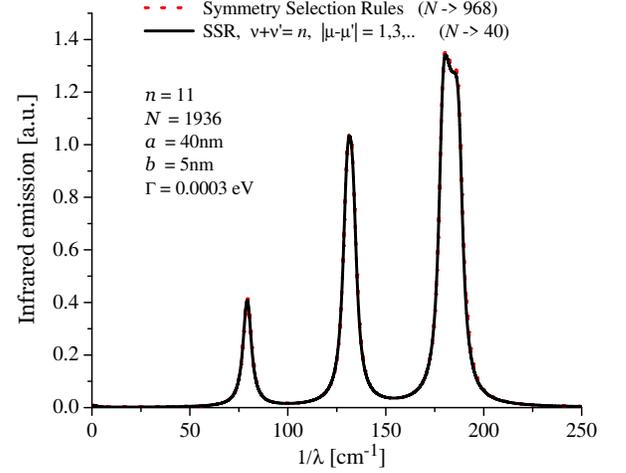}
\caption{Theoretical calculations of the IR spectrum  with LOR (black) and without (red) for the $(GaAs/Al_{x}Ga_{1-x}As)$$^n$ SL with $x=0.3$, valley width $a=40$nm and barrier width $b=5$nm, studied in Ref. [\onlinecite{Helm1991}].  .}\label{IR400w50b1}
\end{figure}
\begin{figure}[h]
\includegraphics [width=220pt]{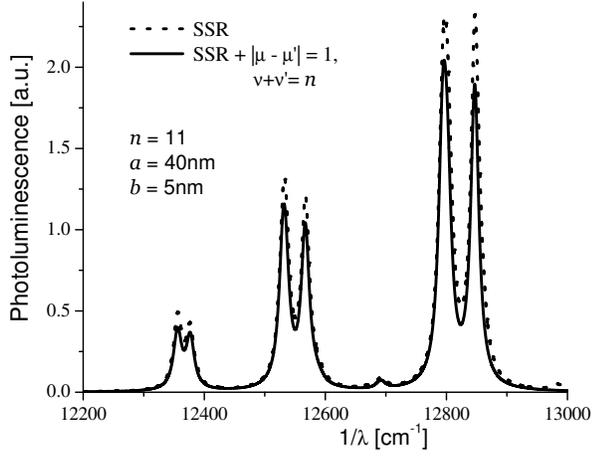}
\caption{Theoretical calculations of the PL spectrum  with LOR (continuous) and without (dotted) for the  $(GaAs/Al_{x}Ga_{1-x}As)$$^n$ SL with $x=0.3$, valley width $a=40$nm and barrier width $b=5$nm, studied in Ref. [\onlinecite{Helm1991}]. .}\label{PL400w50b}
\end{figure}

Among the large amount of results reported in the literature for
the IR spectrum of the $(GaAs/Al_{x}Ga_{1-x}As)$$^n$ SL, let us
consider the SL with $x=0.3$, valley width $a=40$nm and barrier
width $b=5$nm, studied in Ref. [\onlinecite{Helm1991}]. The length
of this SL was 6$\mu$m, which means $n\simeq$133. If we consider a
SL with this number of unit cells, the number of allowed IR
transitions will be about 140,000, for $n_c$=4. But, as mentioned
above,  the same physics comes out when the number of unit cells
is, say, of the order of 10. It is worth recalling that since the
actual SL in Ref. [\onlinecite{Helm1991}] was highly impurified,
additional peaks appeared in the experimental results. Because of
the large well width in these samples, the number of subbands is
large, and the subband positions and widths are extremely
sensitive to potential parameters and effective masses. Assuming a
cladding layer with $x=0.45$ and parameters from Ref.
[\onlinecite{EffectiveMassesFromIoffeInst}], we calculated
accurately the relevant $n$ of the $n+1$ energy eigenvalues and
eigenfunctions for each of the first four subbands, both in the CB
and the VB. For the IR spectra, due to optical transitions in the
CB, the number of  allowed transitions by the SSR when $n$=11 and
$n_c$=4 IS $968$. However, when the LOR are taken into account,
the number of evaluations reduces to $(n-1)(n_c/2)^2=$40. As can
be seen in  figure \ref{IR400w50b1}, there is practically no
difference between the spectrum (red curve) with only SSR and the
spectrum (black curve) where the SSR and LOR are taken into
account. Similar results are found for PL transitions in this SL.
In figure \ref{PL400w50b} we plot the PL spectra for transitions
from the first four subbands in the CB to the first four {\it
heavy hole} subbands. The transitions allowed by  SSR only are
plotted with a dotted curve, while those encompassing to SSR and LOR
are plotted with a continuous curve.
\begin{figure}[h]
\includegraphics [width=220pt]{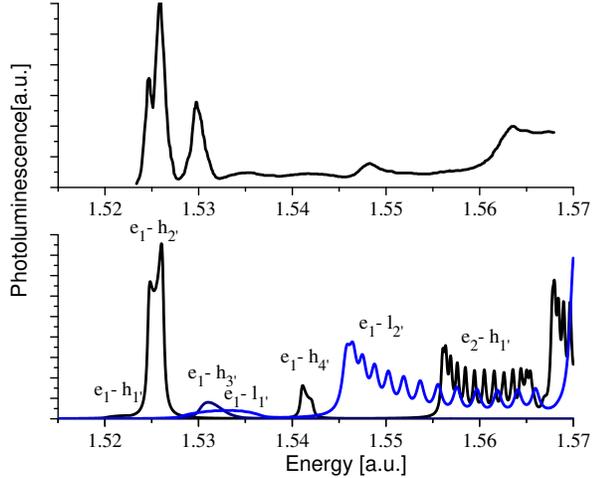}
\caption{PL spectra for  the $(GaAs/Al_{x}Ga_{1-x}As)$$^n$ SL
reported in Ref. [\onlinecite{Fu1989}], with $x=0.3$, valley width
$a=11.8$nm, barrier width $b=1.9$nm, and conduction-band offset
coefficient $Q=0.89$. The upper panel shows the experimental results.
The lower panel shows the theoretical calculation for a SL with
$n$=12 and excitons in the ground state with binding energy of
-5meV.}\label{FuFig5}
\end{figure}

As the third example let us now consider the
$(GaAs/Al_{x}Ga_{1-x}As)$$^n$ SL studied in Ref.
[\onlinecite{Fu1989}], with $x=0.3$, valley width $a=11.8$nm,
barrier width $b=1.9$nm, and conduction-band offset coefficient
$Q=0.89$. The experimental PL spectrum of this sample is shown in
the upper panel of figure \ref{FuFig5}. For this system we
considered an even number of unit cells,  $n$=12.  Thus, the
allowed optical transitions occur whenever $P[\mu+\nu] \neq
P[\mu'+\nu']$. To plot the PL spectrum in the lower panel  of
figure \ref{FuFig5}, we calculated the  energy eigenvalues and
eigenfunctions for all subbands, (four) in the CB, five
$hh$-subbands and two $lh$-subbands in the VB, which implied the
evaluation of 2,366 transition matrix elements.

\begin{figure}[h]
\includegraphics [width=220pt]{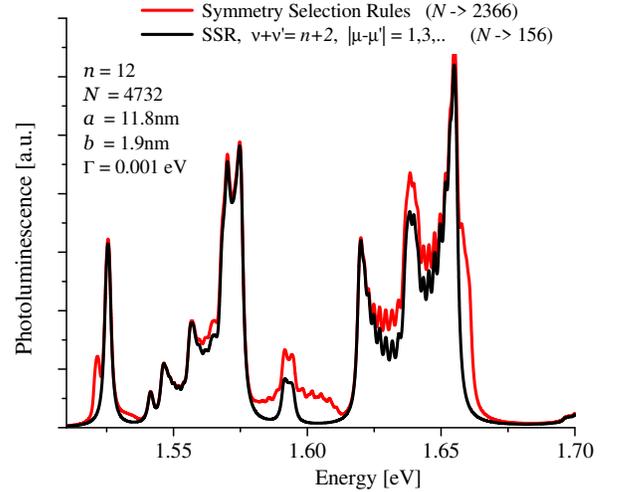}
\caption{The joint PL for the  $e-hh$ and $e-lh$ transitions is plotted for $\Gamma$=1meV, for the $(GaAs/Al_{x}Ga_{1-x}As)$$^n$ SL studied in Ref. [\onlinecite{Fu1989}], with $x=0.3$, valley width $a=11.8$nm, barrier width $b=1.9$nm, and conduction-band offset coefficient $Q=0.89$. The red curve corresponds to a PL that comes out when  all  transitions, allowed by the SSR, are taken into account, and the black curve is obtained when both the SSR and LOR are taken into account.}\label{Fig22}
\end{figure}

The peaks of transitions $e_{1,\nu}$-$h_{1',\nu'}$ and $e_{1,\nu}$-$h_{3',\nu'}$, enlarged in the graph,  are negligible. We aligned the transitions $e_{1,\nu}$-$h_{2',\nu'}$ with the lowest energy peak of the experimental curve, assuming a binding energy of 5meV. To plot this PL spectrum, consider only the contributions of excitons in the ground state and  $\Gamma$=0.1meV.  Again we want to use this example to show that essentially the same PL spectrum can be evaluated with a much smaller number of matrix elements than those required by the SSR.  Above 1.545eV the  resonant structure of the optical transitions  $e_{1,\nu}$-$l_{2',\nu'}$ and $e_{2,\nu}$-$h_{1',\nu'}$ shows that only a small number of  transitions, of the order of the number of unit cells, contribute. For larger $n$ or larger $\Gamma$ the resonant structure  becomes a continuous structure. This is what we have  in figures \ref{Fig22} and \ref{Fig23}, where the joint PL for the  $e-hh$ and $e-lh$ transitions is plotted for $\Gamma$=1meV. In the red curve of figure  \ref{Fig22} we have the PL that comes out when  all  transitions, allowed by the SSR, are taken into account, and the black curve is obtained when both the SSR and LOR are taken into account. The energy fluctuation width, in both cases, is $\Gamma$=1meV. The agreement  is rather good and the reduction in the number of transition matrix elements is enormous.

\section{The Surface and Edge States Rule}\label{SESR}

\begin{figure}[h]
\includegraphics [width=220pt]{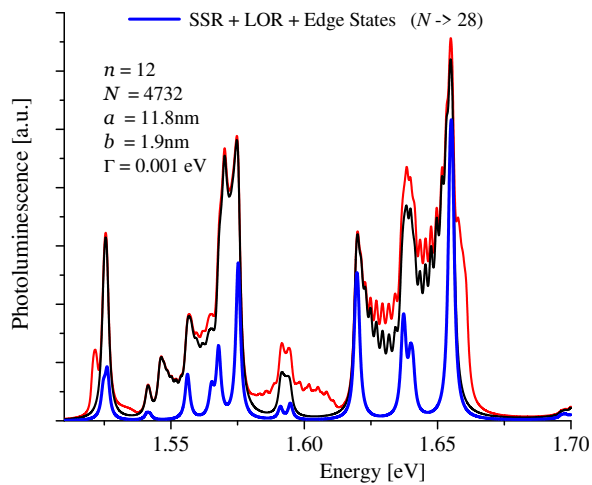}
\caption{Same as figure \ref{Fig22} but with the addition of the PL curve (blue) obtained when we take into account also the surface and edge states rule. }\label{Fig23}
\end{figure}

Just for completeness, we would like to comment briefly on a last but not less important rule, the surface and edge states rule (SESR). In the aim of simplifying the evaluation  of the IR and PL spectra,  we can also stress the relevance that the surface and subband-edge states have in the optical transitions. To make evident this relevance  we copy in Table 2 the matrix elements $\langle {\mu',\nu'}|\frac{\partial}{\partial z}|{\mu,\nu}\rangle$, prior to multiplication by the normalization constants, that were used to evaluate the PL spectrum of the full curve in figure \ref{Fig22}. The darkened matrix elements, evaluated with the edge states, are the leading terms. If we are looking for the qualitative  behavior of the IR or PL spectra, we can use just these matrix elements. This oversimplifying rule  makes contact and coincides, only partially, with the selection  rule of the standard theory. In figure \ref{Fig23} we plot together all the PL curves, obtained when the SSR (red), the SSR and LOR (black)  and the SSR, LOR plus the surface and edge states rule (blue) are taken into account. With these examples it is clear that besides the symmetry selection rules we can use other rules according with the degree of accuracy of the experimental results or the theoretical predictions.

Finally it is worth mentioning that in our theoretical calculations, the change in the effective masses from one layer to the next has always been taken into account and the continuity conditions at the transition points were imposed always on the wave functions and their derivatives. In the last case we did not multiply  by the inverse of the effective masses, as in the Ben Daniel-Duke model.\cite{BenDaniel} It is known\cite{Bastard} that this can easily be done changing $k \rightarrow k/m_{ea}^*$ and $q \rightarrow q/m_{eb}^*$, etc., in the transfer matrices. The impact on the results is negligible.\cite{OnBenDaniel}

\vspace{0.15in}
\setlength{\extrarowheight}{0.0cm}
\begin{center}\small
\begin{tabular}{c l c c c c c}
\multicolumn{7}{c}{\it Table 2: {Transition Matrix Elements} $ \langle {\mu',\nu'}|\frac{\partial}{\partial z}|{\mu,\nu}\rangle$} \\
\multicolumn{7}{c} { with $\nu+\nu'$=n=11 }\\[0.05in]\hline \hline \\[-0.1in] &
& \!\! $\mu'$=2
\!\!&\!\!  $\mu'$=4 &&$\mu'$=1&$\mu'$=3 \\ \;$\mu$ \;&\;$\nu,\,\,\nu' $&&& \; $\mu$\; &&\\
\hline \\[-0.3in]
\\   &  $\;1, \,10 $  & {\bf-41995.5} &-{\bf 5601.99}&&{\bf 43147.6} &{\bf -20794.0}
\\[0.02in]    &   $\;2,\,\, 9 $  & -13935.8&-2066.66&&13931.3&-7179.92
\\[0.02in]    &   $\;3,\,\, 8 $  & -8676.7&-1369.07&&8463.33 &-4588.9
\\[0.02in]    &   $\;4,\,\, 7 $  & -7040.9&-1151.78&&6719.47 &-3786.61
\\[0.02in]   \; 1 \; & $\;5, \,\,6 $  & -6731.5&-1125.58&\;2\;&6295.95 &-3660.88
\\[0.02in]    &   $\;6, \,\,5 $  &-7360.6 &-1247.52&&6755.95 &-4033.15
\\[0.02in]    &   $\;7, \,\,4 $  &-9248.4 &-1580.38&&8343.75&-5091.88
\\[0.02in]    &   $\;8, \,\,3 $  & -13874.8&-2382.3&&12331.1 &-7659.47
\\[0.02in]    &   $\;9, \,\,2 $  & -27581.2&-4747.98&&24221.2 &-15242.8
\\[0.02in]    &   $10,\, 1 $  &{\bf -102213.0} &{\bf -17617.4}&&{\bf 89060.6}&{\bf -56499.7}
\\[0.02in] \hline \\[-0.1in]   &   $\;1, \,10 $  &{\bf22008.4}&{\bf-11059.7}&&{\bf 7468.78}&{\bf 12219.1}
\\[0.02in]    &  $\;2, \,\,9 $  &7218.41&-4086.48&&2398.25&4139.31
\\[0.02in]    &  $\;3, \,\,8 $  &4419.54 &-2714.31&&1445.49&2575.1
\\[0.02in]    &  $\;4, \,\,7 $  &3513.77 &-2291.02&&1136.29&2056.49
\\[0.02in]   \; 3 \;  & $\;5,\,\, 6 $  &3283.2 &-2246.14&\;4\;&1051.96&1916.54
\\[0.02in]    & $\;6, \,\,5 $  &3503.86 &-2495.57&&1112.97&2030.67
\\[0.02in]    & $\;7,\,\, 4 $  & 4296.45&-3164.7&&1352.57&2464.63
\\[0.02in]    &  $\;8, \,\,3 $  &6299.55 &-4767.19&&1964.46&3570.95
\\[0.02in]    &  $\;9, \,\,2 $  &12281.3&-9481.14&&3794.31&6880.33
\\[0.02in]    &  $10, \,1 $  &{\bf 44906.8}&{\bf -35097.2}&&{\bf 13773.1}&{\bf 24931.8}
\\[0.02in] \hline \hline
\end{tabular}\label{Tabla 2}\end{center}
\setlength{\extrarowheight}{0.1cm}

\section{conclusions}

In this paper  additional applications of the
theory of finite periodic systems have been presented and   new selection rules for the evaluation of PL and IR spectra of optoelectronic
devices, whose active region is a superlattice with an arbitrary
number of unit cells, reported. We made clear the
fundamental differences between the standard approach and the
present theory, and consequently the differences in their
prediction capabilities. We have shown that  PL spectra of high
accuracy experiments, that could not be explained before, are now
fully understood. We have shown also that the optical transition
observed experimentally, but forbidden in the standard approach,
are perfectly possible within this theory.  We have shown that
besides the symmetry selection rules, related to the spatial
symmetry of the SL eigenfunctions, one can also introduce  the eigenfunctions' symmetries related
with the intra-subband index, which are behind the leading order selection
rules that make possible a reduction by orders of magnitude in
the number of evaluations of the transition matrix elements,
giving  rise practically to the same PL and IR spectra. With this
theory and the level of accuracy that one can reach, we not only
improve the existing theory but open up the possibility of
discussing other important questions related, in particular, with
the parameters such as the exciton binding energies in SLs, the
effective masses, the band offsets, etc., which define the optical
transitions. This theory can be useful to determine more
accurately their numerical values. At the end, and just for
completeness, we commented on the surface and edge states
selection rule, which together with the SSR and the LOR, imply the
highest reduction in the number of evaluations of the transition
matrix elements. The agreement with the experimental results are
extremely good. 
\section{acknowledgement}

The author acknowledges comments and corrections of H. P. Simanjuntak, A. Robledo-Martinez and J. Grabinsky.

\appendix
\begin{figure}[t]
\includegraphics [width=230pt]{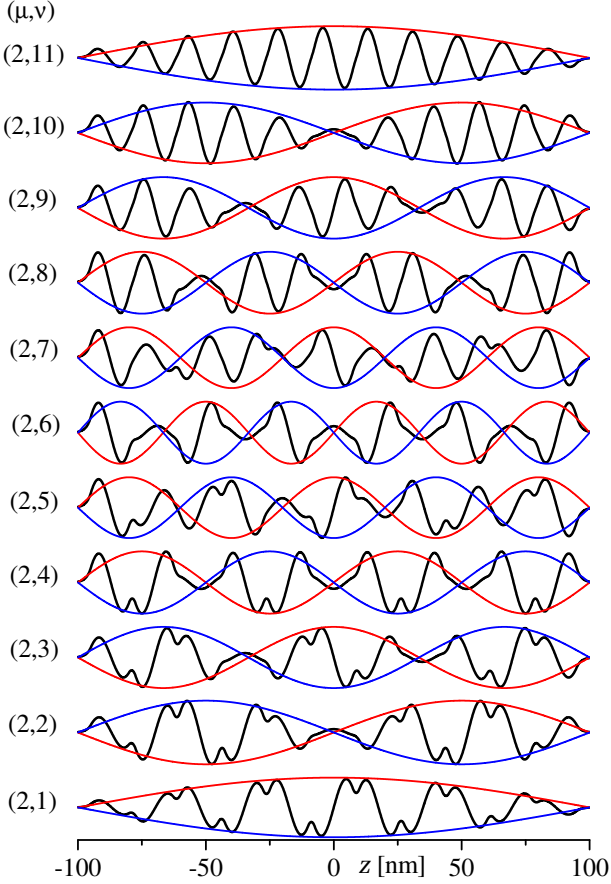}
\caption{Eigenfunctions in second subband of the CB. Besides the spatial symmetry with respect to the origin $z$=0, we have another symmetry with respect to the center of the subband. Thus the eigenfunctions with indices ($\mu,\nu$) and ($\mu,n$+2-$\nu$) have the same symmetry. The envelopes help to visualize these symmetries. }\label{Fig24}
\end{figure}
\section{Eigenfunction symmetries related with the intra-subband index.}

\begin{figure}[h]
\includegraphics [width=230pt]{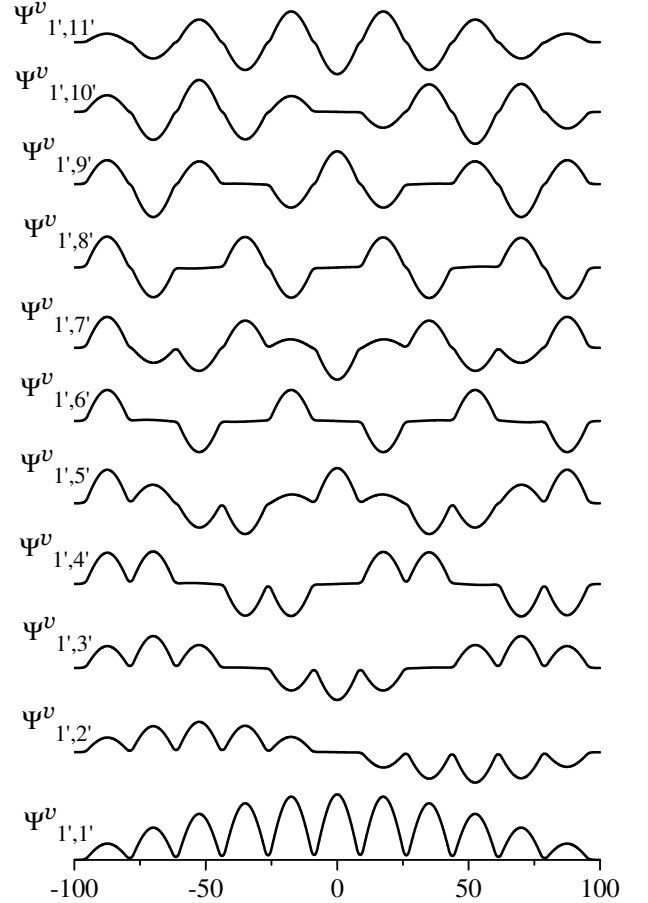}
\caption{Eigenfunctions in the first subband of the VB of the SL considered in figure \ref{Fig24}.  The eigenfunctions with indices ($\mu',\nu'$) and ($\mu',n$+2-$\nu'$) have the same symmetry.}\label{Fig25}
\end{figure}

In addition to the parity symmetries of the SL eigenfunctions with
respect to the middle point of the superlattice, the eigenfunctions
possess another symmetry that can be recognized when we plot the
whole set of intra-subband eigenfunctions. In figures \ref{Fig24}
and \ref{Fig25} we plot all the eigenfunctions in the second
subband of the CB and in the first subband of the VB, for the SL
$(Al_{0.25}Ga_{0.75}As\backslash GaAs)^{n}\backslash
Al_{0.25}Ga_{0.75}As$  with $a$=21nm, $b$=10nm and $n$=10,
considered in section 2 with the PL spectra of figure
\ref{PLMasselink}.

To visualize the symmetry relations between the SL eigenfunctions
of the same subband, and the role of the intra-subband index $\nu$, we
plot in figure \ref{Fig24} guide lines (that look like envelope
functions), together with the eigenfunctions $\{\Psi_{2,\nu}\}$.
With the help of these guide lines we can easily recognize the
symmetries that exist between pairs of  intra-subband
eigenfunctions.  See, for example: the eigenfunctions with indices
$(2,1)$ and $(2,11)$; the eigenfunctions with indices $(2,2)$ and
$(2,10)$;... etc. The same happens when we plot the set of
eigenfunctions of any other subband, in the conduction or valence
band. In figure  \ref{Fig25} we plot the heavy hole eigenfunctions
$\{\Psi_{1',\nu'}\}$ of  the first subband, with similar symmetry
relations between them.  In this system the surface states do not
detach, therefore, as mentioned in section 3, the eigenfunctions
$\Psi_{\mu,\nu}$ and $\Psi_{\mu,n+2-\nu}$ share the same symmetry,
which is then reflected when the transition matrix elements are
evaluated.  As mentioned before,  these symmetries are behind the
matrix-elements values, thus, behind the leading order selection
rules, studied in section 3.

\section{The binding energy of asymmetric excitons}
The calculation of the binding energy for excitons in quantum wells, double quantum wells
and SLs, as
well as impurities effects on the energy spectra, have been of
great interest. Appropriate fitting of the optical-resonances
positions led to study this problem, which is simple to visualize
but rather cumbersome to obtain a rigorous solution. The actual
asymmetry of the confining potential led to search variational
calculations and different models of pseudo-spherical excitons. A
large number of papers, based on the fractional-dimension
approach, were published and binding energies of the exciton in a
quantum well were obtained as functions of the well width. In these
approaches the effective dimension D of the pseudo-spherical
exciton lies between 2 and 3, implying binding energies between 4
and 1 Rydberg. Exciton binding energies in a SL was also addressed
by Pereira et al.,\cite{Pereira1990} using a
variational approach and a trial wave function, with similar
results. For excitons in the $Al_{0.3}Ga_{0.7}As/GaAs$ SL with
valley width $a$=15nm and barrier width $b$=2.5nm they find
$E_{Bhh}\simeq$1.45$E_R$ and $E_{Blh}\simeq$=1.15$E_R$, with $E_R$
the effective Rydberg energy. The same year, Leavitt and Little\cite{Leavitt} published also a simple
variational method for the calculation of binding energies in
quantum confined semiconductor structures. In this model the
expectation energy $\langle w(\zeta) \rangle_0$ for a Hamiltonian
with a Coulomb potential $1/\sqrt{u^2+\zeta^2}$, where $u$ and
$\zeta$ are the dimensionless  in-plane $\rho_e-\rho_{h}$ and
perpendicular $z_e-z_{h}$ electron-hole distances, has been
calculated using a trial wave function that was written in terms
of a variational parameter $\lambda$, defined in such a way that
correct results are obtained for both $\zeta=0$ and $\zeta>>u$, in
the ground state $2k+1+m=1$, i.e. for $k$=0 and $m$=0.

In the following subsection, we extend this model to obtain, besides the ground state expectation energy, the first excited state expectation energy $\langle w(\zeta) \rangle_1$

\subsection{The first excited exciton state in the Leavitt-Little model}
\begin{figure}[h]
\includegraphics [width=220pt]{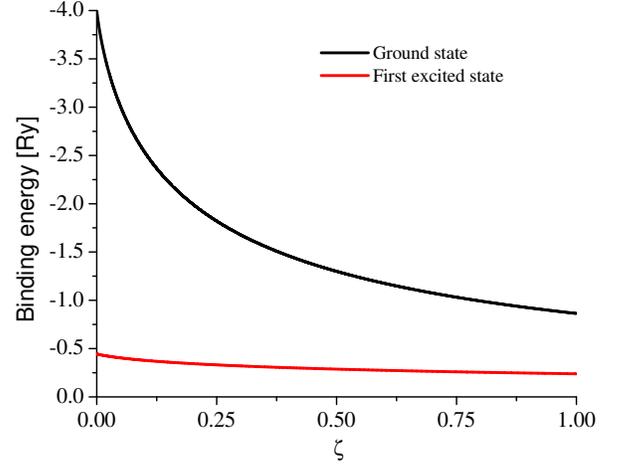}
\caption{The expected exciton binding energies $\langle
w\rangle_0$ (upper curve) and $\langle w\rangle_1$ as FUNCTIONS of
$\zeta$. The difference between the ground and the first excited
state, measured in units of $R_y$, reduces as the exciton
dimension grows from $D$=2 to $D$=3. }\label{fig11to21}
\end{figure}
When the quasi-two-dimensional Schr\"odinger equation  (7), is written in the center of mass and relative coordinates, the radial part, for the azimuthal quantum number $m$=0, is
\begin{eqnarray}
\Bigl[\frac{\hbar^2}{2\mu_\parallel}\frac{1}{\rho}\frac{d}{d\rho}\rho\frac{d}{d\rho}\!-\!\frac{e^2}{\epsilon
\sqrt{\rho^2+z^2}}\Bigr]g_{\eta}(\rho,z)\!=\!E^{(2D)}_{\eta}g_{\eta}(\rho,z)\nonumber\\
\end{eqnarray}
where   $z$=$z_e$-$z_h$ and
$\mu_\parallel$ the in-plane relative {\it e-h} effective mass.
Defining $u$=$\rho/a_0$, $\zeta$=$(z_e$-$z_h)/a_0$,
$w_{\eta}(\zeta)$=$E^(2D)_{\eta}/E_0$ and $G_{\eta}(u;\zeta)$=$a_0
g_{\eta}(u;\zeta)$, where $a_0$=$\epsilon \hbar^2/\mu_\parallel e^2$
and $E_0$=$\mu_\parallel e^4/2\epsilon^2\hbar^2$, we have the
dimensionless differential equation
\begin{eqnarray}
\Bigl(\frac{1}{u}\frac{d}{d u}u\frac{d}{d u}\!-\!\frac{2}{ \sqrt{u^2+\zeta^2}}\Bigr)G_{\eta}(u,\zeta)\!=\!w_{\eta}(\zeta)g_{\eta}(u,\zeta).\nonumber \\
\end{eqnarray}
Knowing the exact solutions: $G_0(u;0)=(1/u)e^{-2u}$ with $w_0(0)=4$, for $\zeta=0$, and $G_0(u;\zeta>>1)=\sqrt{2} e^{-u^2/2\zeta^{3/2}}/u\zeta^{3}$ with $w_0(\zeta)=2/\zeta-2/\zeta^{3/2}$, for $\zeta>>u$, the trial function
\begin{eqnarray}
G_{0}(u,\zeta)=Ne^{-\lambda_0 (\sqrt{u^2+\zeta^2}-\zeta)}
\end{eqnarray}
has been suggested, with $N$ a normalization constant and
$\lambda_0$ a variational parameter chosen as
$\lambda_0=2/(1+2\sqrt{\zeta})$, in such a way that as $\zeta
\rightarrow 0$ ( $\lambda_0 \rightarrow 2$), and  for $\zeta>>1$
($\lambda_0 \rightarrow 1/\sqrt{\zeta}$), the trial wave function
approaches  the correct results. Since the first excited
energy and the exact solutions are also known in these limits:
$w_k=4/(2k+1+2m)^2$ for $\zeta=0$ and
$w_k=2/\zeta(1-(2k+1+m)/\sqrt{\zeta})$ for $\zeta>>u$, we propose
the trial wave function
\begin{eqnarray}
G_{k}(u,\zeta)=Ne^{-\lambda_k (\sqrt{u^2+\zeta^2}-\zeta)}\bigl[ 1-2\lambda_k(\sqrt{u^2+\zeta^2}-\zeta) \bigr]^k \nonumber \\
\end{eqnarray}
with
\begin{eqnarray}
\lambda_k=\frac{2}{2k+1+m+2\sqrt{\zeta}}.
\end{eqnarray}
The expectation values of the binding energy
\begin{eqnarray}
\langle w_k(\zeta)\rangle=\frac{\int_0^\infty G_{k}^*(u,\zeta){\hat{H}_{2D} }G_{k}(u,\zeta) u du}{\int_0^\infty |G_{k}(u,\zeta)|^2 u du},
\end{eqnarray}
become, for $k=0$ and $k=1$,, respectively,
\begin{eqnarray}
\langle w_0(\zeta)\rangle=-\lambda_0^2+\frac{4 \lambda_0+4\zeta^2\lambda_0^4}{1+2\sqrt{\zeta}}e^{2 \zeta \lambda_0} \Gamma(0,2\zeta\lambda_0),
\end{eqnarray}
and
\begin{eqnarray}
\langle w_1(\zeta)\rangle&=&-4 e^{2 \zeta \lambda_1}\zeta^2\lambda_1^4 (3+2 \zeta\lambda_1)\Gamma(0,2\zeta\lambda_1)\nonumber \cr &&+\frac{ 4\lambda_1-\lambda_1^2\bigl(3+2\zeta\lambda_1(5+2\zeta\lambda_1(5+2\zeta\lambda_1))\bigr)}{3+2\zeta\lambda_1}.
\end{eqnarray}
In figure A1 we plot the expectation value of the $hh$-exciton binding energies in the ground and first exited states. The ratio between these energies changes from a factor of 9 in the 2D limit to a factor that tends to 1 for large $\zeta$.


\begin{references}
\bibitem{EMA}   We consider these Hamiltonians in the effective mass approximation, which as Altarelli\cite{AltarelliLesHuches} states, has proven to be simple, accurate and versatile for the calculation of electronic levels, and superior to other methods with the remarkable feature that all reference to the microscopic structure of the host semiconductor is condensed in the effective mass $m^*$
 and the band edge $E_n(0)$.
\bibitem{WhyNotChargePolariz} It is known that in the zincblende structure $E_{g,GaN}=3.2$eV, while in the wurtzite structure  $E_{g,GaN}=3.4$eV. A difference of 0.2eV could not be explained taking into account the charge polarization effect. This effect, as was shown in Fatna an Kunold, might produce a shift in the subband positions of the order of 40meV, as much.
\bibitem{reductions IR} The reductions depend on whether the  number of subbands $n_c$ are even or odd. If  $\nu+\nu'=n$ the reduction is from $(n+1)^2n_c^2/2$ to $(n-1)(n_c/2)^2$ for $n_c$ even and $(n-1)n_c(n_c+1)/2$ for $n_c$ odd. When $\nu+\nu'=n+2$ the reduction is from $(n+1)^2n_c^2/2$ to $(n+1)(n_c/2)^2$ for $n_c$ even and $(n+1)n_c(n_c+1)/2$ for $n_c$ odd.
\bibitem{OnBenDaniel}The differences in the band structure taking into account a) the continuity of the wave functions and their derivatives and the continuity of the wave functions and their derivatives multiplied by the inverse of the effective masses, the so-called  Ben Daniel-Duke continuity conditions, is in the low energy subbands of the order of 0.1meV while in the higher energy subbands of the order of 1meV, as was shown by Bastard\cite{Bastard}
\end{references}
\end{document}